\documentclass[aps,prl,twocolumn,showpacs,floatfix,superscriptaddress,noshowpacs]{revtex4-1}
\usepackage{graphicx}
\usepackage{amsmath}
\usepackage{amssymb}
\usepackage{color}
\usepackage{float}
\usepackage[colorlinks=true,citecolor=blue,linkcolor=blue]{hyperref}

\begin{document}

\flushbottom

\title{Disordered hyperuniformity in superconducting vortex lattices} 

\author{Jos\'e Benito Llorens}
\affiliation{Laboratorio de Bajas Temperaturas y Altos Campos Magn\'eticos (Unidad Asociada UAM-CSIC), Departamento de F\'isica de la Materia Condensada, Instituto Nicol\'as Cabrera and Condensed Matter Physics Center (IFIMAC), Universidad Aut\'onoma de Madrid, E-28049 Madrid,
Spain}

\author{Isabel Guillam\'on}
\affiliation{Laboratorio de Bajas Temperaturas y Altos Campos Magn\'eticos (Unidad Asociada UAM-CSIC), Departamento de F\'isica de la Materia Condensada, Instituto Nicol\'as Cabrera and Condensed Matter Physics Center (IFIMAC), Universidad Aut\'onoma de Madrid, E-28049 Madrid,
Spain}

\author{Ismael Garc\'ia-Serrano}
\affiliation{Laboratorio de Microscopias Avanzadas (LMA) and Instituto de Nanociencia de Arag\'on (INA), Universidad de Zaragoza, E-50018 Zaragoza, Spain}
\affiliation{Instituto de Ciencia de Materiales de Arag\'on (ICMA), Universidad de Zaragoza-CSIC, E-50009 Zaragoza, Spain and Departamento de F\'isica de la Materia Condensada, Universidad de Zaragoza, E-500009 Zaragoza, Spain}

\author{Rosa C\'ordoba}
\affiliation{Laboratorio de Microscopias Avanzadas (LMA) and Instituto de Nanociencia de Arag\'on (INA), Universidad de Zaragoza, E-50018 Zaragoza, Spain}
\affiliation{Instituto de Ciencia de Materiales de Arag\'on (ICMA), Universidad de Zaragoza-CSIC, E-50009 Zaragoza, Spain and Departamento de F\'isica de la Materia Condensada, Universidad de Zaragoza, E-500009 Zaragoza, Spain}
\affiliation{Instituto de Ciencia Molecular, Universidad de Valencia, Catedr\'atico Jos\'e Beltr\'an 2, 46980 Paterna, Spain}

\author{Javier Ses\'e}
\affiliation{Laboratorio de Microscopias Avanzadas (LMA) and Instituto de Nanociencia de Arag\'on (INA), Universidad de Zaragoza, E-50018 Zaragoza, Spain}
\affiliation{Instituto de Ciencia de Materiales de Arag\'on (ICMA), Universidad de Zaragoza-CSIC, E-50009 Zaragoza, Spain and Departamento de F\'isica de la Materia Condensada, Universidad de Zaragoza, E-500009 Zaragoza, Spain}

\author{Jos\'e Mar\'ia De Teresa}
\affiliation{Laboratorio de Microscopias Avanzadas (LMA) and Instituto de Nanociencia de Arag\'on (INA), Universidad de Zaragoza, E-50018 Zaragoza, Spain}
\affiliation{Instituto de Ciencia de Materiales de Arag\'on (ICMA), Universidad de Zaragoza-CSIC, E-50009 Zaragoza, Spain and Departamento de F\'isica de la Materia Condensada, Universidad de Zaragoza, E-500009 Zaragoza, Spain}

\author{M. Ricardo Ibarra}
\affiliation{Laboratorio de Microscopias Avanzadas (LMA) and Instituto de Nanociencia de Arag\'on (INA), Universidad de Zaragoza, E-50018 Zaragoza, Spain}
\affiliation{Instituto de Ciencia de Materiales de Arag\'on (ICMA), Universidad de Zaragoza-CSIC, E-50009 Zaragoza, Spain and Departamento de F\'isica de la Materia Condensada, Universidad de Zaragoza, E-500009 Zaragoza, Spain}

\author{Sebasti\'an Vieira}
\affiliation{Laboratorio de Bajas Temperaturas y Altos Campos Magn\'eticos (Unidad Asociada UAM-CSIC), Departamento de F\'isica de la Materia Condensada, Instituto Nicol\'as Cabrera and Condensed Matter Physics Center (IFIMAC), Universidad Aut\'onoma de Madrid, E-28049 Madrid,
Spain}

\author{Miguel Ortu\~no}
\affiliation{Departamento de F\'isica, CIOyN, Universidad de Murcia, Murcia 30071, Spain}

\author{Hermann Suderow}
\affiliation{Laboratorio de Bajas Temperaturas y Altos Campos Magn\'eticos (Unidad Asociada UAM-CSIC), Departamento de F\'isica de la Materia Condensada, Instituto Nicol\'as Cabrera and Condensed Matter Physics Center (IFIMAC), Universidad Aut\'onoma de Madrid, E-28049 Madrid,
Spain}

\begin{abstract}
Particles occupying sites of a random lattice present density fluctuations at all length scales. It has been proposed that increasing interparticle interactions reduces long range density fluctuations, deviating from random behaviour. This leads to power laws in the structure factor and the number variance that can be used to characterize deviations from randomness which eventually lead to a new form of disorder named disordered hyperuniformity. It is however not yet fully clear how to link density fluctuations with interactions in a disordered hyperuniform system. In superconducting vortex lattices, intervortex interactions are very sensitive to vortex pinning, to the crystal structure of the superconductor and to the value of the magnetic field. This creates lattices with different degrees of disorder. Here we study disordered vortex lattices in several superconducting compounds (Co-doped NbSe$_2$, LiFeAs and CaKFe$_4$As$_4$) and in two amorphous W-based thin films, one with strong nanostructured pinning (W-film-1) and another one with weak or nearly absent pinning (W-film-2). We calculate for each case the structure factor and number variance and compare to calculations on an interacting set of partially pinned particles. We find that interactions manifest either in the presence of small hexagonal bundles (Co-doped NbSe$_2$ and W-film-1), in orientational correlations (W-film-2) or in locking to the crystal lattice (LiFeAs). Random density fluctuations appear when pinning overcomes interactions (W-film-2 close to H$_{c2}$ and LiFeAs at large magnetic fields, as well as in CaKFe$_4$As$_4$). Thus, we conclude that the suppression of density fluctuations in disordered lattices is indeed correlated to the presence of interactions. Furthermore, we find that can describe all pinned vortex lattices within a single framework consisting of a continous deviation from hyperuniformity towards random distributions when increasing the strength of pinning with respect to the intervortex interaction. 
\end{abstract}

\maketitle

\section{Introduction}

\begin{figure}
	\includegraphics[width=0.45\textwidth]{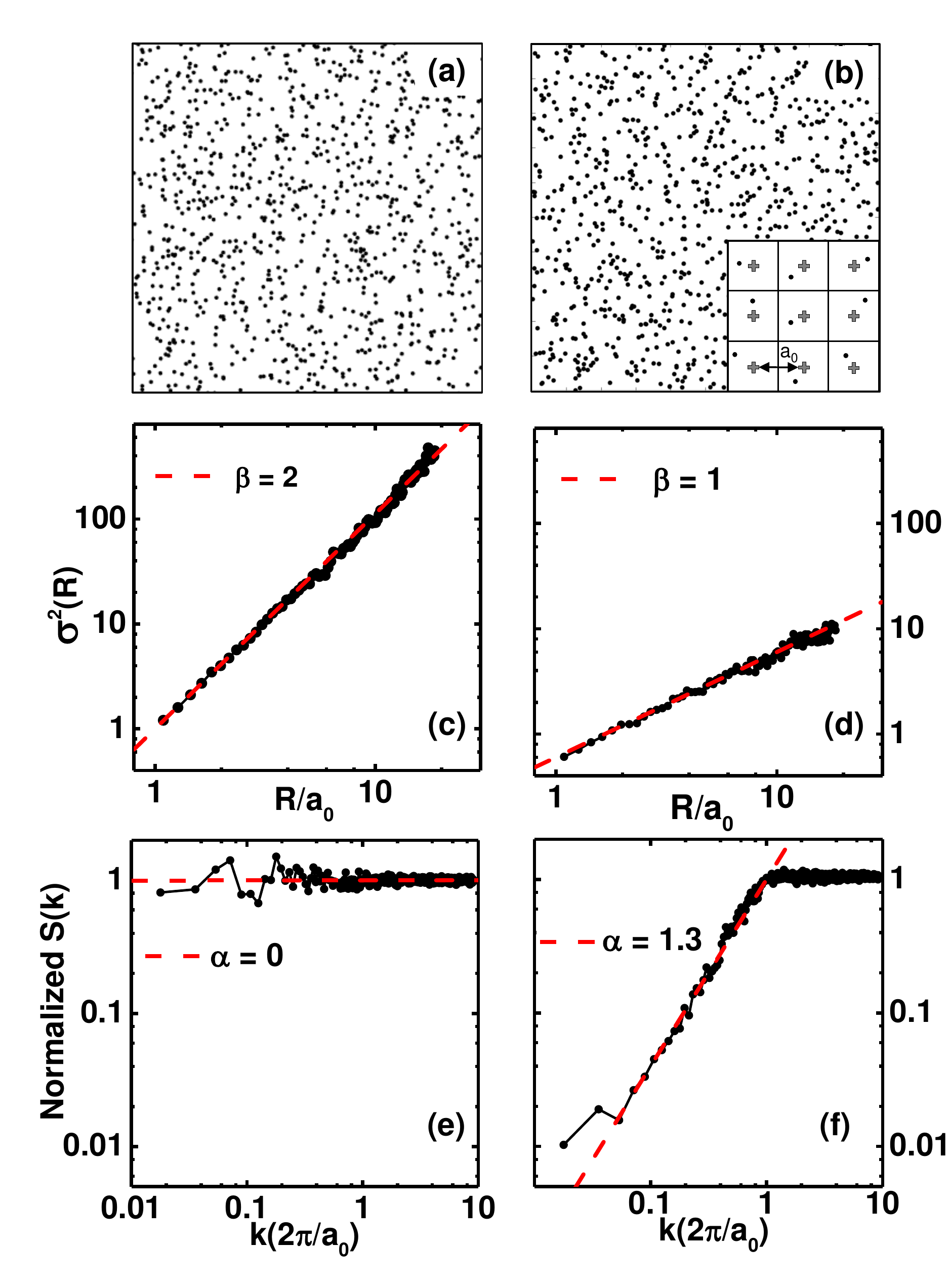}
	\caption{Set of points distributed in a 2D space, obtained by assigning some pixels the value 1 following a random distribution (a) and a hyperuniform distribution (b). Note that both distributions of points are fully disordered. In the lower right inset of (b) we describe schematically how we created the hyperuniform distribution. We started from a square lattice (grey crosses) with lattice constant $a_0$ and added to each lattice point a vector with random coordinates $\vec{r}=(x,y)$ and magnitude $r<2a_0$. This gives the distribution of black dots. (c,d) Variance of the distribution, $\sigma^2$, as a function of the radius normalized to $a_0$, calculated as explained in the text. $\sigma^2(R)$ follows a power law with the exponent $\beta$ (red dotted line). (c) shows the result for the random distribution shown in (a) and (d) for the hyperuniform distribution shown in (b). (e,f) Structure factor $S(k)$ normalized to one for large values of the reciprocal space vector $k$ for (e), the random distribution shown in (a) and (f), the hyperuniform distribution shown in (b). $S(k)$ also follows a power law with exponent $\alpha$ (red dotted lines in e and f). }
	\label{Figure_1_Scheme}
	\end{figure}

Hyperuniformity is a property of a lattice which consists of the absence of density fluctuations at large distances. All ordered lattices, including Moir\'e patterns and quasicrystals are hyperuniform\cite{PhysRevE.68.041113,TORQUATO20181}. Disordered lattices, however, are usually random. A fundamental property of a random distribution is that there are density fluctuations at all length scales. It has been shown that disordered hyperuniform lattices can be created by designing disordered patterns without density fluctuations at large length scales\cite{PhysRevE.68.041113,TORQUATO20181,Man15886,PhysRevE.89.022721,PhysRevE.84.030401,PhysRevE.91.012302,PhysRevLett.121.115501,PhysRevLett.123.068003,PhysRevLett.106.178001,PhysRevLett.115.108301,Man15886,PhysRevB.96.094516,PhysRevB.95.075303}. Recently, hyperuniform behavior has been found in the contact-number between subsystems of particles\cite{PhysRevLett.121.115501,PhysRevLett.123.068003}, although the origin of the suppressed density fluctuations remains under debate\cite{HyperunifJCC}.

Here we address the behavior of vortices in type II superconductors. Vortices are a lattice of whirlpools of currents, each carrying a flux quantum\cite{RevModPhys.66.1125,Brandt_1995}. Vortices repel each other and often form hexagonal and square lattices with intervortex distances $a_0\propto \frac{1}{\sqrt{H}}$. Intervortex interactions are usually repulsive and screened above the penetration depth $\lambda$, which is most often much larger than the intervortex distance $a_0$ \cite{doi:10.1063/1.1753804,RevModPhys.36.45,PhysRevB.24.1572,Buzdin2013}. In thin films,  $\lambda$ strongly increases with decreasing thickness and vortices interact through their stray field, which leads to a long range Coulomb-like $\frac{1}{r}$ interaction \cite{doi:10.1063/1.1754056,PhysRevB.48.6699,PhysRevB.96.184502}. The question we address here is to what extent intervortex interactions can eliminate density fluctuations while allowing the formation of a disordered lattice due to pinning. We analyze vortex lattices with different amounts of disorder in bulk superconductors and in thin films. We make calculations of the positions of particles in a long range interaction potential in presence of disorder. We find a gradual appearance of density fluctuations with increasing disorder.

\section{Methods}

Fig.\,\ref{Figure_1_Scheme} (a,b) shows matrices of points with, respectively, random and hyperuniform disorder. Both images are square and have the same size, with area $A$, and the same number of points $N$. In a random distribution (Fig.\,\ref{Figure_1_Scheme} (a)), there are density fluctuations at all length scales. In a hyperuniform distribution (Fig.\,\ref{Figure_1_Scheme} (b)), density fluctuations disappear at large length scales. The hyperuniform distribution of Fig.\,\ref{Figure_1_Scheme}(b) has been generated by starting with points arranged in a lattice of constant $a_0=\sqrt{A/N}$. This defines a lattice of squares of size $a_0\times a_0$. We then add to each lattice point a vector $\vec{r}$ with random coordinates, whose absolute value $r<2a_0$ (Fig.\,\ref{Figure_1_Scheme}(b)). The result is a random distribution which is spatially uniform for length scales larger than the intercell distance. It is important to realize that there are no further signatures of the ordered lattice in Fig.\,\ref{Figure_1_Scheme}(b) other than the large scale uniformity.

The calculation of the number variance is very useful to discuss density fluctuations\cite{PhysRevE.68.041113,TORQUATO20181,PhysRevB.96.094516}. It is given by $\sigma^2(R)=\langle N^2(R)\rangle-\langle N(R)\rangle^2$, where $N(R)$ is the number of points inside a circle of radius $R$. In the Appendix we illustrate how to calculate $\sigma^2(R)$ using a vortex lattice image. The variance $\sigma^2(R)$ increases with $R$ as a power law $\sigma^2(R) \propto R^{\beta}$. In  In a random distribution of points (Fig.\,\ref{Figure_1_Scheme}(a)), $\sigma^2(R)$  grows as the dimension, (i.e with the area, according to the large number law) so that, $\beta=2$ (Fig.\,\ref{Figure_1_Scheme}(c)). On the other hand, in a hyperuniform distribution (Fig.\,\ref{Figure_1_Scheme}(b)), $\sigma^2(R)$  grows as the dimension minus one (i.e with the perimeter) with $\beta = 1$ (Fig 1(d)). In lattices with orientational order, $\sigma^2(R)$ shows a dip each time $R$ is somewhat smaller than integer multiples of the average interparticle distance\cite{klatt2020cloaking}.

To calculate the structure factor $S(k)$ we use the Fourier transform of the image of the vortex positions and make the radial average over the Fourier transform. In the Fourier space, the structure factor $S(k)$ shows a Bragg peak at $a_0$ and decreases at small $k$  with the power law $S(k) \propto k^{\alpha}$.  In the random distribution $\alpha = 0$ (Fig.\,\ref{Figure_1_Scheme}(e)), while in the hyperuniform distribution $\alpha >1$ (Fig.\,\ref{Figure_1_Scheme}(f)). There is a relation between the exponents of $S(k)$ and those of $\sigma^2(R)$. When $0 < \alpha < 1$, then $\beta = 2- \alpha$ and when $\alpha >1$, $\beta$ remains locked at 1.  

We chose vortex lattice images with a considerable degree of disorder obtained in Refs.\cite{Guillamon2014,PhysRevB.78.174518,PhysRevB.85.214505,PhysRevB.97.134501,GUILLAMON201470}. Notice that studying the prototypical pure 2H-NbSe$_2$ or other materials with well ordered lattices would make little sense in the context of this work\cite{PhysRevLett.62.214,PhysRevLett.101.166407,RevModPhys.79.353,Suderow_2014,Troyanovski1999}. We calculate $S(k)$ and $\sigma^2(R)$ and discuss the power laws as a function of $k$ and $R$. We discuss data on a nanostructured W-based thin film (W-film-1) with strong pinning and polycrystalline vortex lattice arrangements that disorder at high magnetic fields. We show results in Co-doped 2H-NbSe$_2$, a system with strong point-like pinning centers that also leads to polycrystalline hexagonal lattices which increase the level of disorder when increasing magnetic fields\cite{PhysRevB.78.174518}. We discuss a W-based thin film with weak 1D disorder potential and perfectly ordered vortex lattices that disorders at high magnetic fields (W-film-2)\cite{Guillamon2014}. Finally, we discuss two iron based superconductors with strongly disordered lattices at all magnetic fields, LiFeAs, CaKFe$_4$As$_4$\cite{PhysRevB.85.214505,PhysRevB.97.134501}.

The results in W-film-1 have not been published. We acquire the image in zero field cooled conditions at 100 mK using the system described in Refs.\,\cite{Guillamon2014,GUILLAMON201470,doi:10.1063/1.3567008}. The sample has been made using a focused ion beam assisted deposition and has a composition similar to the composition of W-film-2, which has perfectly ordered lattices in a large range of magnetic fields, described in Ref.\,\cite{Guillamon2014}. The critical temperature is of 5 K\cite{Guillam_n_2008,Cordoba2013}. However, contrary to W-film-2, here the substrate has strong random thickness modulations at a length scale of about 200 nm, which considerably enhance pinning. The vortex lattice rearranges accordingly, showing a polycrystalline pattern, which we discuss in the Appendix (Fig.\,\ref{Figure_5_Bundles}).

\section{Results}

\begin{figure}
	\includegraphics[width=0.4\textwidth]{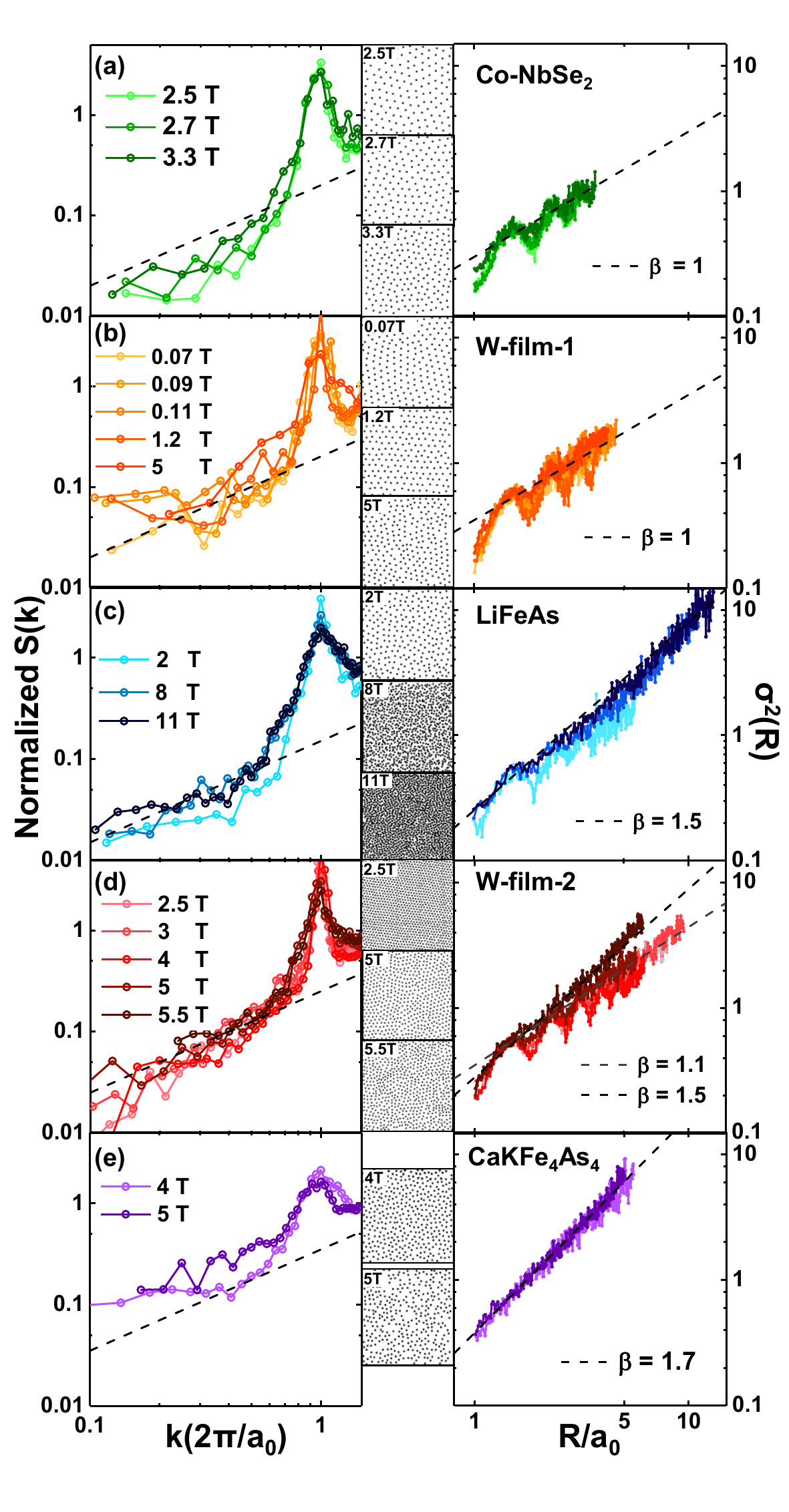}
	\caption{In the left column, we show the structure factor $S(k)$ obtained for vortex lattices in many different materials as a function of the reciprocal lattice vector $k$ in units of $\frac{2\pi}{a_0}$ in each image. In the right column, we show the variance $\sigma^2(R)$ obtained in the same materials. A few images of vortex positions are shown between graphs. In (a) we show results in Co-doped NbSe$_2$ with data taken from Ref.\cite{PhysRevB.78.174518}. The lateral size of the images is of 375 nm. In (b) we show results obtained in a W-based thin film with strong pinning (W-film-1). The lateral size of the images is of 1700 nm, 1000 nm and 750 nm. In (c) we show results in LiFeAs, with data from Ref.\cite{PhysRevB.85.214505}. The lateral size of the images is of 500 nm. In (d) we show results in a highly ordered W-based thin film (W-film-2) with a very weak 1D disorder potential, data from Ref.\cite{Guillamon2014}. The lateral size of the images is of 1000 nm (2.5 T and 3 T), 550 nm (4 T, 5 T) and 500 nm (5.5 T). In (e) we show results in pure CaKFe$_4$As$_4$, from Ref.\cite{PhysRevB.97.134501}. The lateral size of the images is of 400 nm and 470 nm. Points are joined by lines as a guide. We plot all data in log scales and provide the power law dependencies with exponents $\alpha=1$ for $S(k)$ and $\beta$ as shown in the legends of the figures for $\sigma^2(R)$. See also Table \ref{TableResults}.}
	\label{Figure_3_Results}
	\end{figure}

In Fig.\,\ref{Figure_3_Results}(a,b) we show results for vortex lattices consisting of patches of hexagonal lattices, i.e.\ polycrystalline arrangements of varying sizes. The vortex lattice positions are shown in the central panels of Fig.\,\ref{Figure_3_Results}(a,b). In Fig.\,\ref{Figure_3_Results}(a) we show results in Co-doped 2H-NbSe$_2$\cite{PhysRevB.78.174518}.  We find $S(k)\propto k^{\alpha}$ with $\alpha\geq 1$ and $\beta=1$. In Fig.\,\ref{Figure_3_Results}(b) we show results in W-film-1. We find $\alpha\geq 1$ and $\beta=1$, very similar as in the previous case. Both of these systems show close to hyperuniform behavior. There is short range hexagonal order in all images at length scales well above $a_0$. We can see this in the oscillations appearing in $\sigma^2(R)$ close to integers of $a_0$.

In Fig.\,\ref{Figure_3_Results}(c) we show results in LiFeAs, from Ref.\,\cite{PhysRevB.85.214505}. The vortex lattice is highly disordered above about 2 T, with no clear hexagonal patterns observed at any length scale. As discussed in Ref.\cite{PhysRevB.85.214505}, the structure factor has a square shaped orientational dependence, which shows that overall there is a tendency of the vortex lattice to lock its orientation to the square crystal lattice. Note that oscillations in $\sigma^2(R)$ are much less pronounced than in other cases, although these are clearly visible at 2 T. The coefficient $\alpha$ is slightly smaller than one and $\beta$ is close to one at 2 T but increases with the magnetic field. This situation is close to a disordered hyperuniform arrangement at low magnetic fields. Note that, despite the presence of disorder, the vortex interaction is not negligible since it leads to an orientational locking of the disordered vortex lattice with the crystal lattice.

In Fig.\,\ref{Figure_3_Results}(d) we show results in W-film-2, from Ref.\cite{Guillamon2014}. At small magnetic fields, where hexagonal order is nearly perfect, with a few dislocations that proliferate when increasing the magnetic field. The lattice disorders when increasing the magnetic field above about 4 T and at 5 T, the lattice has no long range positional nor orientational order\cite{Guillamon2014}. $S(k)$ and $\sigma^2(R)$ remain with the same power law dependencies, with $\beta=1.1$ and $\alpha\geq 1$ for magnetic fields below or equal to 5 T. We also see oscillatory dependence of $\sigma^2(R)$, indicating short range orientational correlations. Notice that, although the lattice at 5T has neither long range positional nor orientational order (see Ref.\cite{Guillamon2014}), the length scale for orientational order is sufficiently large to provide $\beta\approx1$, i.e. near to disordered hyperuniform behavior. When reaching 5.5 T, the decay length for orientational order goes from about five times $a_0$ down to a couple of $a_0$\cite{Guillamon2014}. The oscillations in $\sigma^2(R)$ vanish totally at 5.5 T. But there is also a strong deviation from hyperuniformity, with an increase of $\beta$ to $1.6$. Thus, the onset of strong disorder leads to a tendency to form a random distribution of vortices.

Next we analyze highly disordered vortex lattices in CaKFe$_4$As$_4$, from Ref.\,\cite{PhysRevB.97.134501}. We find (Fig.\,\ref{Figure_3_Results}(e)) a considerable deviation from disordered hyperuniform behavior, with $\beta$ close to or larger than $1.5$ and $\alpha$ less than 1. We find no signatures of oscillations in $\sigma^2(R)$.

We summarize all results in Table \ref{TableResults}.

\begin{table}[]
\begin{tabular}{c|ccc}
Material        & $\beta$      & Oscillations in $\sigma^2(R)$ \\ \hline
Co-NbSe$_2$ & 1            & Yes               \\
W-film-1       & 1            & Yes               \\
LiFeAs & 1.3 & Yes-No       \\
W-film-2       & 1.1-1.5     & Yes-No           \\
CaKFe$_4$As$_4$ & 1.7          & No               
\end{tabular}\caption{Exponent $\beta$ of the variance $\sigma^2(R)$ obtained from the data in Fig.\,\ref{Figure_3_Results}. Crystalline order shows the presence or not of a visible oscillating pattern in $\sigma^2(R)$. Random behavior is characterized by $\beta=2$ and no crystalline order. The deviation from random towards disordered hyperuniform is seen by $\beta$ decreasing from $2$ towards $1$ in absence of crystalline order.}\label{TableResults}
\end{table}

\begin{figure*}
	\includegraphics[width=0.85\textwidth]{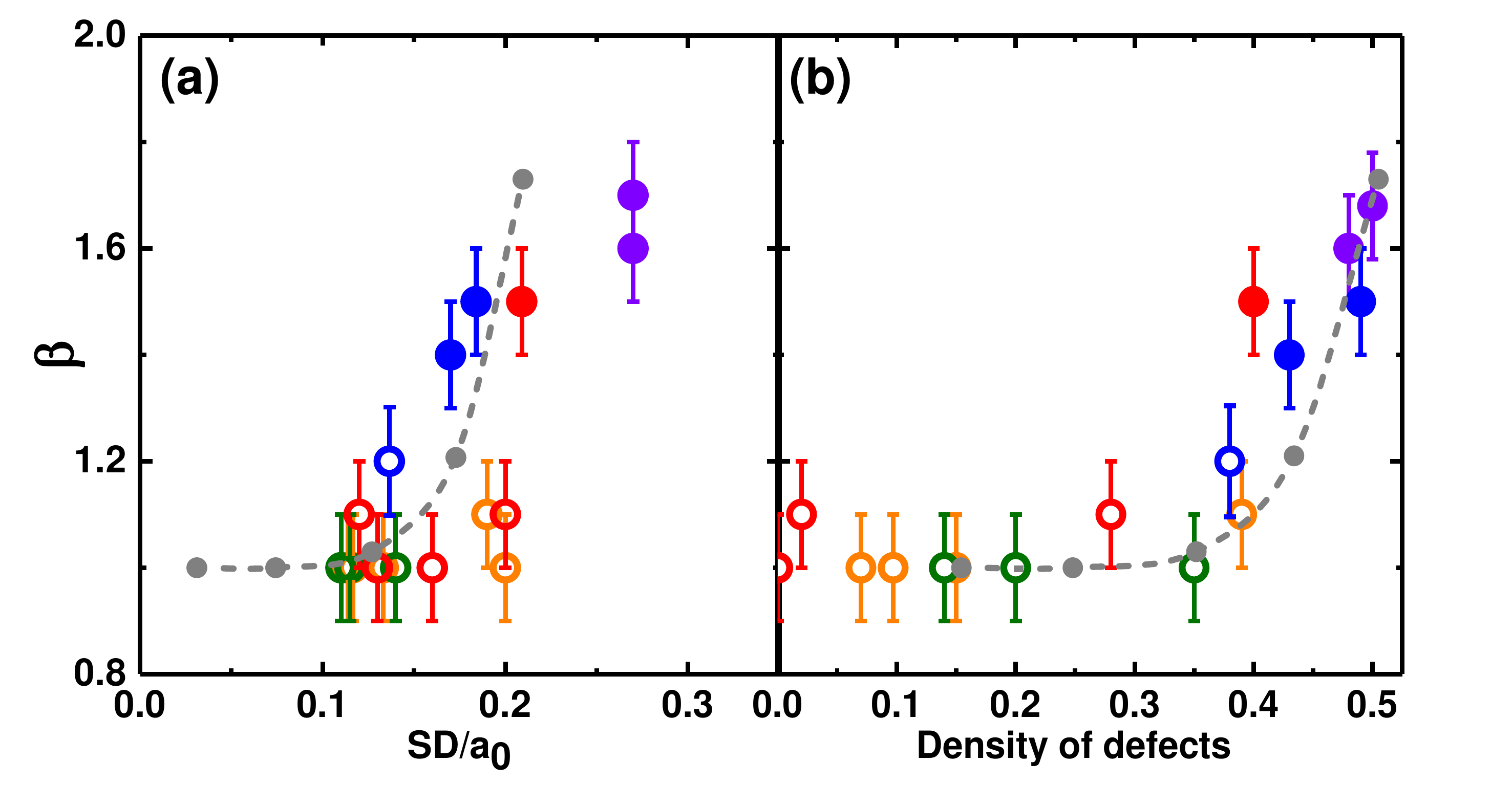}
	\caption{In (a) we show the exponent of $\sigma^2(R)$, $\beta$, as a function of the standard deviation normalized by the intervortex distance $\frac{SD}{a_0}$. In (b) we show $\beta$ as a function of the defect density. We show results (circles) for the systems of Fig.\ref{Figure_3_Results}, with similar colors as in Fig.\ref{Figure_3_Results}. Green is Co-NbSe$_2$, orange is W-film-1, blue is LiFeAs, red is W-film-2 and violet is CaKFe$_4$As$_4$. Open circles indicate results where  $\sigma^2(R)$ has an oscillatory behavior, showing the presence of hexagonal order. Filled circles show situations with a fully disordered vortex lattice and no oscillations in $\sigma^2(R)$. Grey points show the result of the numerical simulation discussed in the text (dashed grey line is a guide to the eye).}
	\label{Figure_4_Beta_Variance}
	\end{figure*}

Using our results, we can plot $\beta$ as a function of the standard deviation (SD) in the nearest neighbor vortex positions normalized to the intervortex distance $a_0$, $\frac{SD}{a_0}$ and the density of defects in the vortex lattice (Fig.\,\ref{Figure_4_Beta_Variance}; we call a defect a vortex with coordination number different than six, see also the Appendix).

If we start from an ordered lattice, we are close to $\beta$=1 and $\frac{SD}{a_0}$ as well as the defect density close to zero. We see that when there are oscillations in $\sigma^2(R)$ (open points in Fig.\,\ref{Figure_4_Beta_Variance}),   $\beta$=1 although the $\frac{SD}{a_0}$ and the defect density can be quite large. Notice that there are no data with $\beta$=1 and $\frac{SD}{a_0}$ larger than about 30-40\% of the intervortex distance. This ressembles a Lindemann criterion. Above a certain fluctuation amplitude ($\frac{SD}{a_0}$), the ordered lattice is unstable.

When we have a randomly disordered vortex lattice, we expect $\beta$ tending towards 2 and large values for $\frac{SD}{a_0}$ and of the defect density. This indeed occurs for the fully disordered lattices of CaKFe$_4$As$_4$, with LiFeAs at 11 T and with W-film-2 at 5.5 T (filled points in Fig.\,\ref{Figure_4_Beta_Variance}). For fully disordered hyperuniform, or close to hyperuniform behaviour, we expect $\beta$ close to 1. Either $\frac{SD}{a_0}$ or the density of defects, or both should be of course large. In LiFeAs at 2 T we observe $\beta$ close to 1, and a small $\frac{SD}{a_0}$ but a large amount of defects. As shown in Ref.\,\cite{PhysRevB.85.214505}, the disordered lattice is locked to the crystal lattice, following its orientation. In W-film-2 at 5 T we observe similar parameters, $\beta$ close to 1, and a small $\frac{SD}{a_0}$ but a large amount of defects. There are oscillations in $\sigma^2(R)$ highlighting that orientational order is maintained up to several $a_0$. Thus, the appearance of disordered hyperuniformity, or the decrease in density fluctuations in disordered lattices, is linked to the presence of interactions, either in the form of locking to the crystal lattice (LiFeAs) or in short range orientational order (W-film-2).

We have made a numerical simulation distributing points randomly on a square sample, with a density similar than the vortex density in the experiments. We use periodic boundary conditions and fix a percentage of points in their positions (between 10\% and 50\%). The rest is free to move and tend to minimize the potential energy of the system. We use a $1/r$ interaction, which converges faster than the usual intervortex interaction and produces a similar distribution. We do not take into account the fact that the intervortex interaction is screened above $\lambda$, although this should not greatly influence the comparison to the experiment. The procedure selects a point at random and chooses possible new positions displaced by a small distance from the original position in a random direction. If the new position minimizes the system energy, the point is moved to the new position. After a number of changes in the positions equal to one thousand times the number of points, we consider that changes are minute and we stop the algorithm. Then, we calculate $S(k)$ and $\sigma^2(R)$, obtaining $\beta$ as discussed previously, for the final minimum energy configuration. We show the result in Fig.\,\ref{Figure_4_Beta_Variance} as grey points.

We find that the model fits particularly well distributions of vortices where there are no maxima in $\sigma^2(R)$ (filled circles in Fig.\,\ref{Figure_4_Beta_Variance}). For relatively small levels of disorder, with a defect density less than 30-40\%, the result is close to hyperuniform behavior. However, a full disordered hyperuniform behavior, with $\beta=1$ and a defect density much larger than 30-40\%, does not occur. Moreover, the model shows a gradual increase of $\beta$ away from hyperuniform behavior when the amount of defects increases. This suggests that the increase in $\beta$ is due to increasing number of defects.

In vortex lattices, the number of defects is the consequence of the balance between the pinning energy and the intervortex interactions. For example, in LiFeAs, disorder appears relatively far from H$_{c2}$, with a vortex lattice that interacts strongly through the crystalline lattice, as shown by the four-fold symmetry in the structure factor discussed in Ref.\,\cite{PhysRevB.85.214505}. This leads to $\beta$ close to 1, with a large amount of defects, although $\frac{SD}{a_0}$ is maintained to relatively small values. When increasing the magnetic field there is a strong tendency towards random behavior. In W-film-2 with $\beta$ close to 1, strong disorder appears at 5 T (which is 0.78H$_{c2}$), although with a smaller defect density and with enough orientational order to present maxima in $\sigma^2(R)$. At a slightly larger magnetic field, at 5.5 T (which is 0.85H$_{c2}$), the lattice of W-film-2 yields to random disorder and $\beta$ increases with a jump to 1.5, with practically the same $\frac{SD}{a_0}$ but a larger amount of defects. 

We note that the model produces lattices that have very small values of $\frac{SD}{a_0}$ at Fig.\,\ref{Figure_4_Beta_Variance}(a). These lattices do not appear in the experiment (no points below $\frac{SD}{a_0}\lesssim$ 0.1). When the vortex lattice still has short range positional order (open dots in Fig.\,\ref{Figure_4_Beta_Variance}), $\beta=1$. In this group of lattices, we have lattices that show defects and polycrystalline arrangements (W-film-1 and Co-NbSe$_2$, shown in the Appendix), as well as ordered lattices with a very small number of defects (open red points of W-film-2 in Fig.\,\ref{Figure_4_Beta_Variance}). In the latter the positional correlations decay exponentially with distance\cite{Guillamon2014}, which explains why the $\frac{SD}{a_0}$ remains above 0.1. Fully ordered hexagonal lattices (or the vortex Bragg glass with algebraically decaying positional correlation) provide $\beta\approx 1$ for near to zero $\frac{SD}{a_0}$ in the representation of Fig.\,\ref{Figure_4_Beta_Variance}\cite{Guillamon2014,Klein2001,PhysRevLett.72.1530,PhysRevB.52.1242,PhysRevB.43.130,PhysRevB.55.626}.

\section{Discussion}

The vortex lattice arrangements are a consequence of the balance between elastic and pinning energies. In Co-doped NbSe$_2$ and in the W-film-1 thin film, pinning is strong but structured, leading to hexagonal vortex clusters observed at all magnetic fields. On the other hand, in CaKFe$_4$As$_4$ pinning centers are so strong and randomly distributed that the vortex lattice is essentially randomly disordered in the whole range of magnetic fields studied \cite{PhysRevB.97.134501,PhysRevMaterials.2.074802,PhysRevB.99.104506,PhysRevB.100.064524,Ishida2019,PhysRevB.99.104506}. In W-film-2, where the disorder potential is very weak, the vortex lattice only disorders when it is very soft, very close to H$_{c2}$. But then, the vortex distribution shows a strong tendency to disorder randomly, because intervortex interactions are very weak. In LiFeAs we have qualitatively the same behavior as in CaKFe$_4$As$_4$, but with weaker pinning. Furthermore, the intervortex interaction with non-local contributions due to the effect of the crystal lattice symmetry\cite{PhysRevB.24.1572,PhysRevB.55.R8693}, is still important and responsible for decreasing density fluctuations.

The vortex lattice of LiFeAs depends strongly on the temperature range where the magnetic field is applied. There are measurements showing hexagonal vortex lattices in the same magnetic field range\cite{PhysRevB.99.161103}, whereas the ones we have used here \cite{PhysRevB.85.214505} and neutron scattering experiments \cite{PhysRevLett.104.187001} provide disordered lattices. Notice that the disordered lattices discussed here are locked to the crystal lattice. This is a rather peculiar combination of long range interaction and disorder. Locking can be explained by nonlocal corrections to the London model that favor a fourfold vortex lattice\cite{PhysRevB.24.1572,PhysRevB.55.R8693}.

It is relevant to note that the only disordered lattices with close to hyperuniform behavior ($\beta\approx 1$, points in Fig.\,\ref{Figure_4_Beta_Variance} corresponding to LiFeAs at small magnetic fields and to W-film-2 at 5 T) have relatively small $\frac{SD}{a_0}$. In the case of LiFeAs, data follow closely the calculations. Therefore we expect that fully disordered vortex lattices will not fall to the behavior of lattices showing fluctuations in $\sigma^2(R)$ as a consequence of short range order (open points with $\beta\approx 1$ in the representation of Fig.\,\ref{Figure_4_Beta_Variance}), but rather follow the smooth increase of $\beta$ with the amount of disorder predicted by the model.

Calculations show that the vortex glass can present disordered hyperuniformity in a range of magnetic field and temperatures in presence of strong repulsive interactions and quenched disorder\cite{PhysRevB.96.094516}. Authors propose a phase diagram with the near to disordered hyperuniform behavior in between the Bragg glass and the random vortex glass. Our results confirm indeed the presence of this intermediate state and show that it can be obtained as a balance between interaction and pinning.

In a recent work, vortex lattices at very small magnetic fields have been analyzed in view of their hyperuniform properties. Authors have analyzed images of the high T$_c$ cuprate superconductor Bi$_2$Sr$_2$CaCu$_2$O$_{8+\delta}$ with magnetic Bitter decoration in presence of disorder\cite{PhysRevResearch.1.033057}. At small magnetic fields, vortices are very far apart and their mutual repulsion is small\cite{Prozorov2008,llorens2019observation}. Vortex arrangements are then strongly influenced by their interaction with pinning centers and their mutual interaction plays a minor role. Furthermore, at high temperatures, close to the transition to the normal state, the vortex lattice melts, leading to the vortex liquid which is a dynamic tangle of vortices\cite{Cubitt1993,Zeldov1995}. Therefore experiments at small magnetic fields are made by cooling from the liquid phase, which results in quenched vortex arrangements\cite{MARCHEVSKY19972083}. Authors of Ref.\cite{PhysRevResearch.1.033057} conclude that long wavelength fluctuations are systematically suppressed in the vortex lattice at small magnetic fields, as a consequence of the hydrodynamic properties of the liquid phase, which leads to deviations from fully random vortex distributions.

\section{Conclusions}

In summary, we have analyzed the conditions for the formation of disordered hyperuniform vortex lattices in superconductors at high magnetic fields. The vortex lattice shows a tendency away from random behavior and towards hyperuniformity when the amount of pinning centers is between 30\% and 40\%. We find that the length scale of the interaction plays a minor role in determining the strength of density fluctuations. Instead, the balance between pinning and vortex lattice stiffness controls density fluctuations, with a continous variation between density fluctuations and disorder in the lattice. We show that the decreased density fluctuations require intervortex interactions. We conclude that we can identify emergent correlations in a vortex lattice using the structure factor $S(k)$ and number variance $\sigma^2(R)$ and establish the disordered hyperuniform vortex lattice, which is a new correlated vortex glass, characterized by decreased density fluctuations.

\section{Acknowledgments}
We are very grateful for Tetsuo Hanaguri for sharing with us data in raw form. We also acknowledge discussions with Charles Reichhardt, who draw our attention to this problem, and with Yanina Fasano and Maria Iavarone. This work was supported by the Spanish State Agency for Research (AEI, FIS2017-84330-R, MAT2017-82970-C2-1-R, MAT2017-82970-C2-2-R, RED2018-102627-T, RYC-2014-15093 and MDM-2014-0377), from the Arag\'on Regional Government (Construyendo Europa desde Arag\'on) through the project E13-17R with European Social Fund funding, from the Comunidad de Madrid through program NANOFRONTMAG-CM (S2013/MIT-2850) and by EU program Cost CA16218 (Nanocohybri). I.G. acknowledges support by the European Research Council PNICTEYES (grant agreement 679080). M.O. acknowledges support by Fundaci\'on S\'eneca grant 19907/GERM/15. We also acknowledge SEGAINVEX at UAM.

\section{Appendix}

\begin{figure*}
	\includegraphics[width=0.95\textwidth]{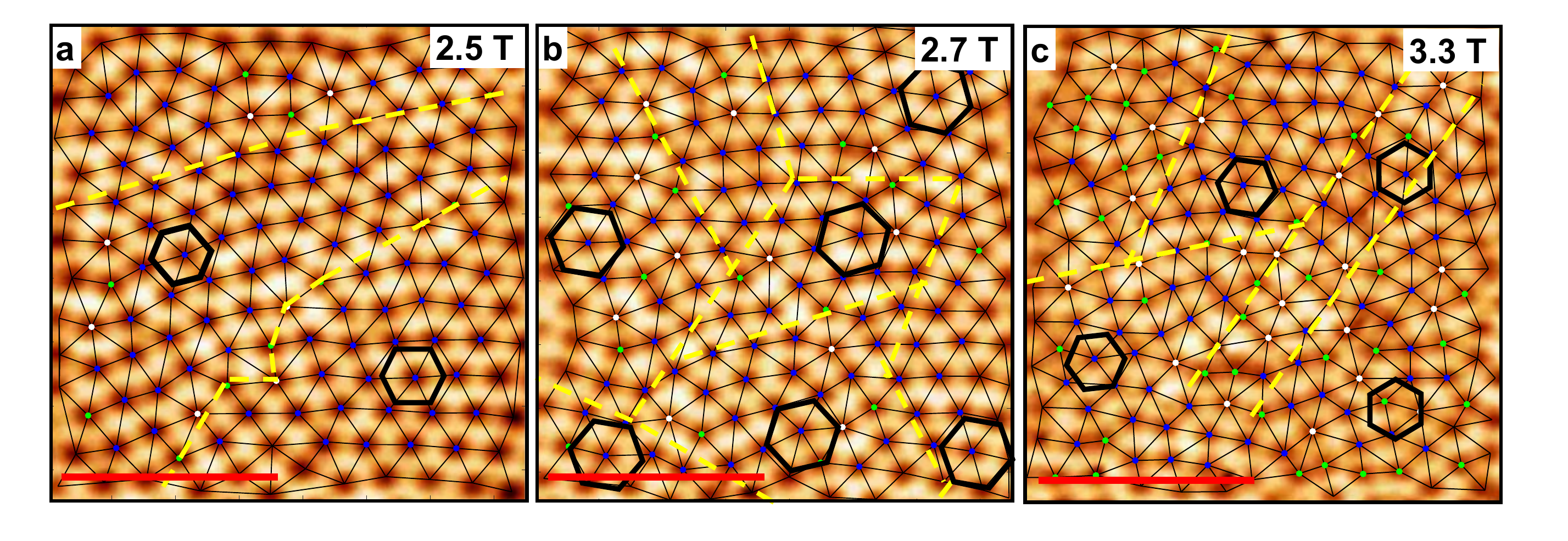}
	\caption{In (a-c) we show results in Co-doped NbSe$_2$, from Ref.\,\cite{PhysRevB.78.174518}. Vortices are shown as black regions. Red bars are 120 nm long. The position of each vortex, identified by the method mentioned in the text, is shown by blue dots. Vortices with less than five nearest neighbors are shown as green dots and with seven nearest neighbors as white dots. Pairs of such vortices provide one dislocation. Black lines provide the Delaunay triangulation of the vortex lattice. Hexagons show the orientation of the lattice in different parts of the images. Lattices showing different orientations are separated by yellow dashed lines.}
	\label{Figure_5_Bundles}
	\end{figure*}

\begin{figure*}
	\includegraphics[width=0.95\textwidth]{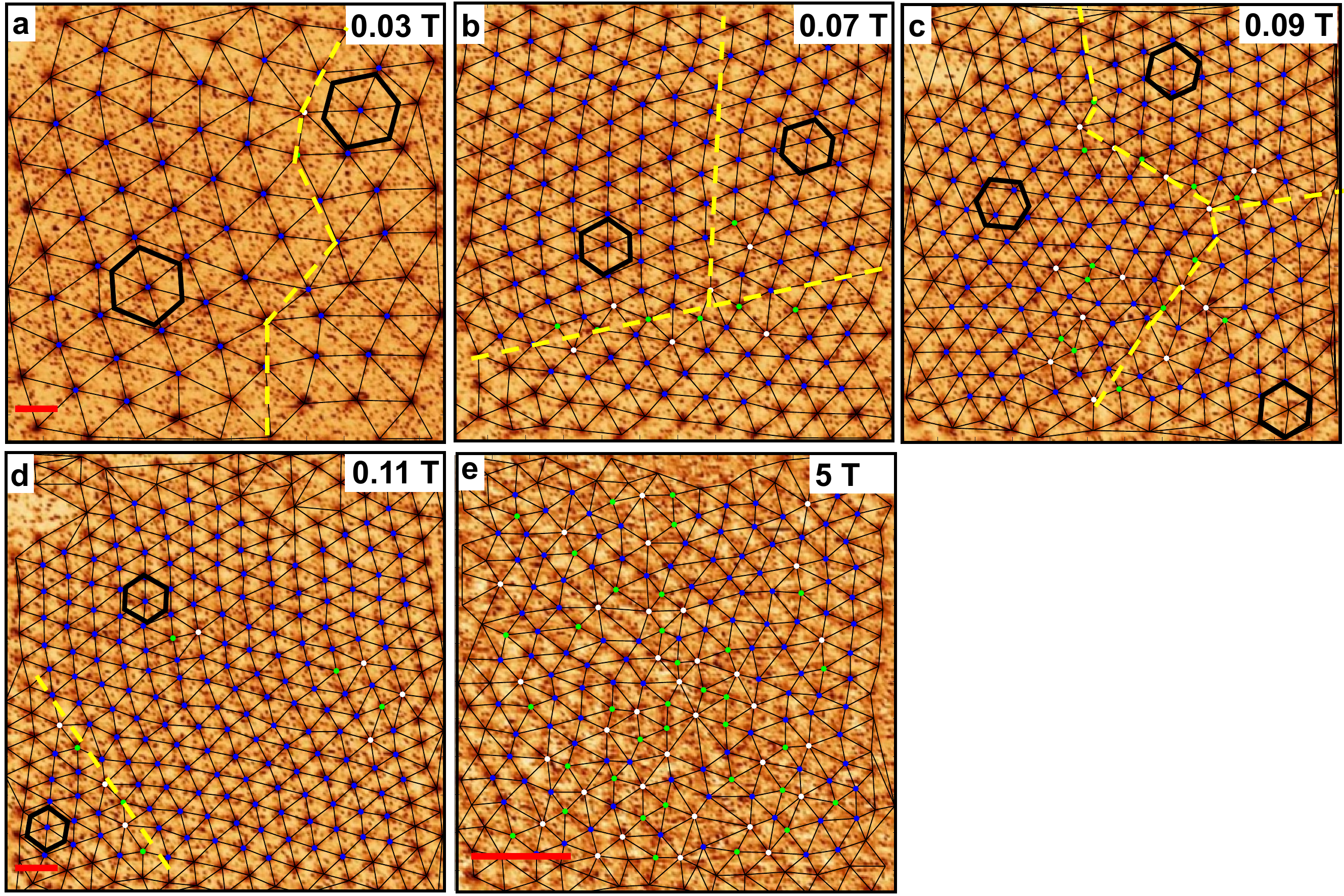}
	\caption{In (a-e) we show the vortex lattice in the W-film-1 thin film for different values of the magnetic field, marked at each figure. We represent vortices, triangulation, lattice orientation and red scale bars as in Fig.\,\ref{Figure_5_Bundles}.}
	\label{Figure_5_W}
	\end{figure*}

In the Fig.\,\ref{Figure_5_Bundles} we show the vortex lattice in Co-doped NbSe$_2$ and in the  Fig.\,\ref{Figure_5_W} the vortex lattice in the W-film-1 thin film with strong pinning. Regions with hexagonal order are observed at all magnetic fields (Fig.\,\ref{Figure_5_Bundles}(a-c)), and the amount of defects can be also quite large, of about 40\%\cite{PhysRevB.78.174518}.

In W-film-1, vortices get pinned by differences in the thickness of the thin film, as described in detail in Ref.\cite{Guillam_n_2008,PinningWbasedStrong}. In the Fig.\,\ref{Figure_5_W}(a-e) we show results with increasing magnetic fields in W-film-1. We see that the vortex lattice remains with regions showing hexagonal order at all magnetic fields. We observe that regions with different vortex lattice orientation are separated by regions with a large amount of dislocations. The smallest ordered regions appear at relatively large magnetic fields. Further increase of the magnetic field towards H$_{c2}$ leads to a disordered random configuration, shown in \ref{Figure_5_W}(e), where we can still identify hexagonal ordered regions of finite size. This lattice is accordingly still hyperuniform, but with a large density of defects (about 40\%).

\begin{figure}
	\includegraphics[width=0.45\textwidth]{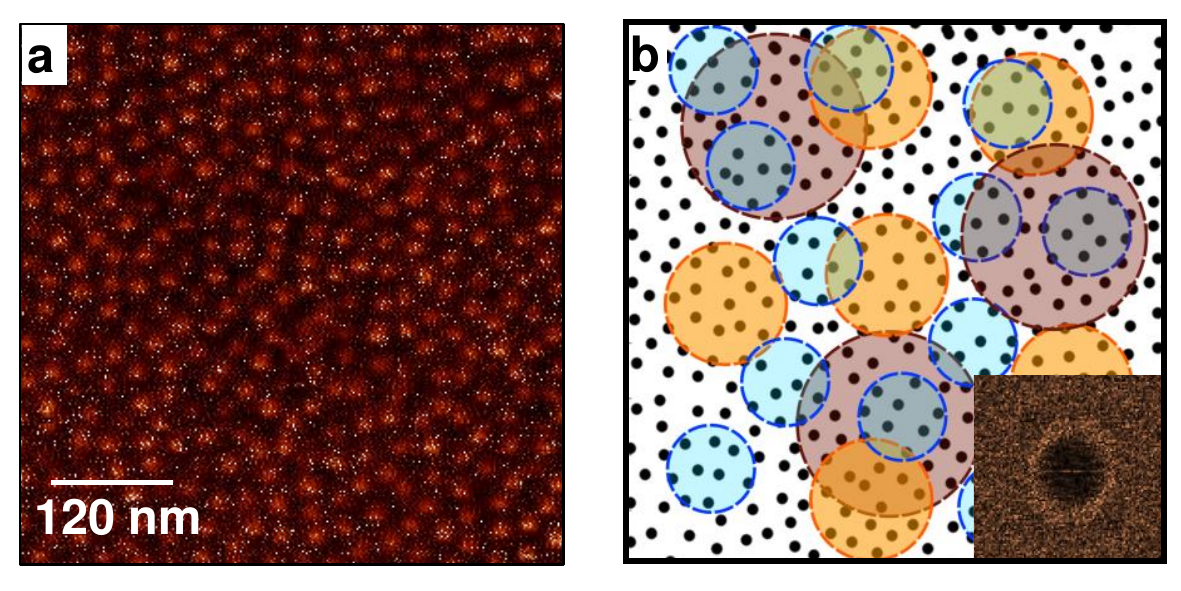}
	\caption{Example for the calculation of $\sigma^2(R)$ and $S(k)$ in a vortex lattice image. In (a) we show the image obtained in CaKFe$_4$As$_4$ at 4 T. Vortices are the white colored patches. The vortex lattice is disordered. In (b) we show as black points the vortex positions of (a). These are obtained by calculating the center of mass of each white patch in (a). In the inset we show the Fourier transform of the image of the vortex positions. To calculate $\sigma^2(R)$, we produce randomly located circles with varying radius $R$, as those shown schematically in different colors. Circles with the same radii are shown in the same color.}
	\label{Figure_2_Example}
	\end{figure}

To calculate $\sigma^2(R)$ and $S(k)$ we start by finding vortex positions following Ref.\cite{Guillamon2014}. We maximize the contrast in the image, inserting a threshold that gives a clear view of vortices as single colored and extended disks. We calculate the center of mass of each disk, and use this to identify the position of each vortex. This leads to the matrix of points shown in Fig.\,\ref{Figure_2_Example}(b). We then make the Fourier transform to find $S(k)$. To calculate $\sigma^2(R)$ we follow Ref.\,\cite{PhysRevB.96.094516}. We generate circles of size $R$ centered at randomly generated positions and increase $R$ from the average intervortex distance $a_0$ to nearly the size of the image. We use two conditions. First, circles have to be complete and within the image. Second, circles cannot overlap. We schematically show a few circles in Fig.\,\ref{Figure_2_Example}(b). In each circle, we count $N(R)$, the number of vortices inside the circle and obtain $\sigma^2(R)=\langle N^2(R)\rangle-\langle N(R)\rangle^2$ by averaging over many circles. When $R$ is small, we obtain in one trial many circles spanning the whole image. When $R$ is large, we obtain just a few circles. We make the calculation in such a way as to increase the number of random tries giving the center of the circles with $R$, taking care that we average over at least one hundred circles for all $R$.

\begin{thebibliography}{55}%
\makeatletter
\providecommand \@ifxundefined [1]{%
 \@ifx{#1\undefined}
}%
\providecommand \@ifnum [1]{%
 \ifnum #1\expandafter \@firstoftwo
 \else \expandafter \@secondoftwo
 \fi
}%
\providecommand \@ifx [1]{%
 \ifx #1\expandafter \@firstoftwo
 \else \expandafter \@secondoftwo
 \fi
}%
\providecommand \natexlab [1]{#1}%
\providecommand \enquote  [1]{``#1''}%
\providecommand \bibnamefont  [1]{#1}%
\providecommand \bibfnamefont [1]{#1}%
\providecommand \citenamefont [1]{#1}%
\providecommand \href@noop [0]{\@secondoftwo}%
\providecommand \href [0]{\begingroup \@sanitize@url \@href}%
\providecommand \@href[1]{\@@startlink{#1}\@@href}%
\providecommand \@@href[1]{\endgroup#1\@@endlink}%
\providecommand \@sanitize@url [0]{\catcode `\\12\catcode `\$12\catcode
  `\&12\catcode `\#12\catcode `\^12\catcode `\_12\catcode `\%12\relax}%
\providecommand \@@startlink[1]{}%
\providecommand \@@endlink[0]{}%
\providecommand \url  [0]{\begingroup\@sanitize@url \@url }%
\providecommand \@url [1]{\endgroup\@href {#1}{\urlprefix }}%
\providecommand \urlprefix  [0]{URL }%
\providecommand \Eprint [0]{\href }%
\providecommand \doibase [0]{http://dx.doi.org/}%
\providecommand \selectlanguage [0]{\@gobble}%
\providecommand \bibinfo  [0]{\@secondoftwo}%
\providecommand \bibfield  [0]{\@secondoftwo}%
\providecommand \translation [1]{[#1]}%
\providecommand \BibitemOpen [0]{}%
\providecommand \bibitemStop [0]{}%
\providecommand \bibitemNoStop [0]{.\EOS\space}%
\providecommand \EOS [0]{\spacefactor3000\relax}%
\providecommand \BibitemShut  [1]{\csname bibitem#1\endcsname}%
\let\auto@bib@innerbib\@empty
\bibitem [{\citenamefont {Torquato}\ and\ \citenamefont
  {Stillinger}(2003)}]{PhysRevE.68.041113}%
  \BibitemOpen
  \bibfield  {author} {\bibinfo {author} {\bibfnamefont {S.}~\bibnamefont
  {Torquato}}\ and\ \bibinfo {author} {\bibfnamefont {F.~H.}\ \bibnamefont
  {Stillinger}},\ }\href {\doibase 10.1103/PhysRevE.68.041113} {\bibfield
  {journal} {\bibinfo  {journal} {Phys. Rev. E}\ }\textbf {\bibinfo {volume}
  {68}},\ \bibinfo {pages} {041113} (\bibinfo {year} {2003})}\BibitemShut
  {NoStop}%
\bibitem [{\citenamefont {Torquato}(2018)}]{TORQUATO20181}%
  \BibitemOpen
  \bibfield  {author} {\bibinfo {author} {\bibfnamefont {S.}~\bibnamefont
  {Torquato}},\ }\href {\doibase https://doi.org/10.1016/j.physrep.2018.03.001}
  {\bibfield  {journal} {\bibinfo  {journal} {Physics Reports}\ }\textbf
  {\bibinfo {volume} {745}},\ \bibinfo {pages} {1 } (\bibinfo {year} {2018})},\
  \bibinfo {note} {hyperuniform States of Matter}\BibitemShut {NoStop}%
\bibitem [{\citenamefont {Man}\ \emph {et~al.}(2013)\citenamefont {Man},
  \citenamefont {Florescu}, \citenamefont {Williamson}, \citenamefont {He},
  \citenamefont {Hashemizad}, \citenamefont {Leung}, \citenamefont {Liner},
  \citenamefont {Torquato}, \citenamefont {Chaikin},\ and\ \citenamefont
  {Steinhardt}}]{Man15886}%
  \BibitemOpen
  \bibfield  {author} {\bibinfo {author} {\bibfnamefont {W.}~\bibnamefont
  {Man}}, \bibinfo {author} {\bibfnamefont {M.}~\bibnamefont {Florescu}},
  \bibinfo {author} {\bibfnamefont {E.~P.}\ \bibnamefont {Williamson}},
  \bibinfo {author} {\bibfnamefont {Y.}~\bibnamefont {He}}, \bibinfo {author}
  {\bibfnamefont {S.~R.}\ \bibnamefont {Hashemizad}}, \bibinfo {author}
  {\bibfnamefont {B.~Y.~C.}\ \bibnamefont {Leung}}, \bibinfo {author}
  {\bibfnamefont {D.~R.}\ \bibnamefont {Liner}}, \bibinfo {author}
  {\bibfnamefont {S.}~\bibnamefont {Torquato}}, \bibinfo {author}
  {\bibfnamefont {P.~M.}\ \bibnamefont {Chaikin}}, \ and\ \bibinfo {author}
  {\bibfnamefont {P.~J.}\ \bibnamefont {Steinhardt}},\ }\href {\doibase
  10.1073/pnas.1307879110} {\bibfield  {journal} {\bibinfo  {journal}
  {Proceedings of the National Academy of Sciences}\ }\textbf {\bibinfo
  {volume} {110}},\ \bibinfo {pages} {15886} (\bibinfo {year} {2013})},\
  \Eprint
  {http://arxiv.org/abs/https://www.pnas.org/content/110/40/15886.full.pdf}
  {https://www.pnas.org/content/110/40/15886.full.pdf} \BibitemShut {NoStop}%
\bibitem [{\citenamefont {Jiao}\ \emph {et~al.}(2014)\citenamefont {Jiao},
  \citenamefont {Lau}, \citenamefont {Hatzikirou}, \citenamefont
  {Meyer-Hermann}, \citenamefont {Corbo},\ and\ \citenamefont
  {Torquato}}]{PhysRevE.89.022721}%
  \BibitemOpen
  \bibfield  {author} {\bibinfo {author} {\bibfnamefont {Y.}~\bibnamefont
  {Jiao}}, \bibinfo {author} {\bibfnamefont {T.}~\bibnamefont {Lau}}, \bibinfo
  {author} {\bibfnamefont {H.}~\bibnamefont {Hatzikirou}}, \bibinfo {author}
  {\bibfnamefont {M.}~\bibnamefont {Meyer-Hermann}}, \bibinfo {author}
  {\bibfnamefont {J.~C.}\ \bibnamefont {Corbo}}, \ and\ \bibinfo {author}
  {\bibfnamefont {S.}~\bibnamefont {Torquato}},\ }\href {\doibase
  10.1103/PhysRevE.89.022721} {\bibfield  {journal} {\bibinfo  {journal} {Phys.
  Rev. E}\ }\textbf {\bibinfo {volume} {89}},\ \bibinfo {pages} {022721}
  (\bibinfo {year} {2014})}\BibitemShut {NoStop}%
\bibitem [{\citenamefont {Kurita}\ and\ \citenamefont
  {Weeks}(2011)}]{PhysRevE.84.030401}%
  \BibitemOpen
  \bibfield  {author} {\bibinfo {author} {\bibfnamefont {R.}~\bibnamefont
  {Kurita}}\ and\ \bibinfo {author} {\bibfnamefont {E.~R.}\ \bibnamefont
  {Weeks}},\ }\href {\doibase 10.1103/PhysRevE.84.030401} {\bibfield  {journal}
  {\bibinfo  {journal} {Phys. Rev. E}\ }\textbf {\bibinfo {volume} {84}},\
  \bibinfo {pages} {030401} (\bibinfo {year} {2011})}\BibitemShut {NoStop}%
\bibitem [{\citenamefont {Dreyfus}\ \emph {et~al.}(2015)\citenamefont
  {Dreyfus}, \citenamefont {Xu}, \citenamefont {Still}, \citenamefont {Hough},
  \citenamefont {Yodh},\ and\ \citenamefont {Torquato}}]{PhysRevE.91.012302}%
  \BibitemOpen
  \bibfield  {author} {\bibinfo {author} {\bibfnamefont {R.}~\bibnamefont
  {Dreyfus}}, \bibinfo {author} {\bibfnamefont {Y.}~\bibnamefont {Xu}},
  \bibinfo {author} {\bibfnamefont {T.}~\bibnamefont {Still}}, \bibinfo
  {author} {\bibfnamefont {L.~A.}\ \bibnamefont {Hough}}, \bibinfo {author}
  {\bibfnamefont {A.~G.}\ \bibnamefont {Yodh}}, \ and\ \bibinfo {author}
  {\bibfnamefont {S.}~\bibnamefont {Torquato}},\ }\href {\doibase
  10.1103/PhysRevE.91.012302} {\bibfield  {journal} {\bibinfo  {journal} {Phys.
  Rev. E}\ }\textbf {\bibinfo {volume} {91}},\ \bibinfo {pages} {012302}
  (\bibinfo {year} {2015})}\BibitemShut {NoStop}%
\bibitem [{\citenamefont {Hexner}\ \emph {et~al.}(2018)\citenamefont {Hexner},
  \citenamefont {Liu},\ and\ \citenamefont {Nagel}}]{PhysRevLett.121.115501}%
  \BibitemOpen
  \bibfield  {author} {\bibinfo {author} {\bibfnamefont {D.}~\bibnamefont
  {Hexner}}, \bibinfo {author} {\bibfnamefont {A.~J.}\ \bibnamefont {Liu}}, \
  and\ \bibinfo {author} {\bibfnamefont {S.~R.}\ \bibnamefont {Nagel}},\ }\href
  {\doibase 10.1103/PhysRevLett.121.115501} {\bibfield  {journal} {\bibinfo
  {journal} {Phys. Rev. Lett.}\ }\textbf {\bibinfo {volume} {121}},\ \bibinfo
  {pages} {115501} (\bibinfo {year} {2018})}\BibitemShut {NoStop}%
\bibitem [{\citenamefont {Hexner}\ \emph {et~al.}(2019)\citenamefont {Hexner},
  \citenamefont {Urbani},\ and\ \citenamefont
  {Zamponi}}]{PhysRevLett.123.068003}%
  \BibitemOpen
  \bibfield  {author} {\bibinfo {author} {\bibfnamefont {D.}~\bibnamefont
  {Hexner}}, \bibinfo {author} {\bibfnamefont {P.}~\bibnamefont {Urbani}}, \
  and\ \bibinfo {author} {\bibfnamefont {F.}~\bibnamefont {Zamponi}},\ }\href
  {\doibase 10.1103/PhysRevLett.123.068003} {\bibfield  {journal} {\bibinfo
  {journal} {Phys. Rev. Lett.}\ }\textbf {\bibinfo {volume} {123}},\ \bibinfo
  {pages} {068003} (\bibinfo {year} {2019})}\BibitemShut {NoStop}%
\bibitem [{\citenamefont {Zachary}\ \emph {et~al.}(2011)\citenamefont
  {Zachary}, \citenamefont {Jiao},\ and\ \citenamefont
  {Torquato}}]{PhysRevLett.106.178001}%
  \BibitemOpen
  \bibfield  {author} {\bibinfo {author} {\bibfnamefont {C.~E.}\ \bibnamefont
  {Zachary}}, \bibinfo {author} {\bibfnamefont {Y.}~\bibnamefont {Jiao}}, \
  and\ \bibinfo {author} {\bibfnamefont {S.}~\bibnamefont {Torquato}},\ }\href
  {\doibase 10.1103/PhysRevLett.106.178001} {\bibfield  {journal} {\bibinfo
  {journal} {Phys. Rev. Lett.}\ }\textbf {\bibinfo {volume} {106}},\ \bibinfo
  {pages} {178001} (\bibinfo {year} {2011})}\BibitemShut {NoStop}%
\bibitem [{\citenamefont {Weijs}\ \emph {et~al.}(2015)\citenamefont {Weijs},
  \citenamefont {Jeanneret}, \citenamefont {Dreyfus},\ and\ \citenamefont
  {Bartolo}}]{PhysRevLett.115.108301}%
  \BibitemOpen
  \bibfield  {author} {\bibinfo {author} {\bibfnamefont {J.~H.}\ \bibnamefont
  {Weijs}}, \bibinfo {author} {\bibfnamefont {R.}~\bibnamefont {Jeanneret}},
  \bibinfo {author} {\bibfnamefont {R.}~\bibnamefont {Dreyfus}}, \ and\
  \bibinfo {author} {\bibfnamefont {D.}~\bibnamefont {Bartolo}},\ }\href
  {\doibase 10.1103/PhysRevLett.115.108301} {\bibfield  {journal} {\bibinfo
  {journal} {Phys. Rev. Lett.}\ }\textbf {\bibinfo {volume} {115}},\ \bibinfo
  {pages} {108301} (\bibinfo {year} {2015})}\BibitemShut {NoStop}%
\bibitem [{\citenamefont {Le~Thien}\ \emph {et~al.}(2017)\citenamefont
  {Le~Thien}, \citenamefont {McDermott}, \citenamefont {Reichhardt},\ and\
  \citenamefont {Reichhardt}}]{PhysRevB.96.094516}%
  \BibitemOpen
  \bibfield  {author} {\bibinfo {author} {\bibfnamefont {Q.}~\bibnamefont
  {Le~Thien}}, \bibinfo {author} {\bibfnamefont {D.}~\bibnamefont {McDermott}},
  \bibinfo {author} {\bibfnamefont {C.~J.~O.}\ \bibnamefont {Reichhardt}}, \
  and\ \bibinfo {author} {\bibfnamefont {C.}~\bibnamefont {Reichhardt}},\
  }\href {\doibase 10.1103/PhysRevB.96.094516} {\bibfield  {journal} {\bibinfo
  {journal} {Phys. Rev. B}\ }\textbf {\bibinfo {volume} {96}},\ \bibinfo
  {pages} {094516} (\bibinfo {year} {2017})}\BibitemShut {NoStop}%
\bibitem [{\citenamefont {Sadovskyy}\ \emph {et~al.}(2017)\citenamefont
  {Sadovskyy}, \citenamefont {Wang}, \citenamefont {Xiao}, \citenamefont
  {Kwok},\ and\ \citenamefont {Glatz}}]{PhysRevB.95.075303}%
  \BibitemOpen
  \bibfield  {author} {\bibinfo {author} {\bibfnamefont {I.~A.}\ \bibnamefont
  {Sadovskyy}}, \bibinfo {author} {\bibfnamefont {Y.~L.}\ \bibnamefont {Wang}},
  \bibinfo {author} {\bibfnamefont {Z.-L.}\ \bibnamefont {Xiao}}, \bibinfo
  {author} {\bibfnamefont {W.-K.}\ \bibnamefont {Kwok}}, \ and\ \bibinfo
  {author} {\bibfnamefont {A.}~\bibnamefont {Glatz}},\ }\href {\doibase
  10.1103/PhysRevB.95.075303} {\bibfield  {journal} {\bibinfo  {journal} {Phys.
  Rev. B}\ }\textbf {\bibinfo {volume} {95}},\ \bibinfo {pages} {075303}
  (\bibinfo {year} {2017})}\BibitemShut {NoStop}%
\bibitem [{\citenamefont {Diamant}(2019)}]{HyperunifJCC}%
  \BibitemOpen
  \bibfield  {author} {\bibinfo {author} {\bibfnamefont {H.}~\bibnamefont
  {Diamant}},\ }\href {\doibase 10.36471/JCCM\_September\_2019\_02} {\bibfield
  {journal} {\bibinfo  {journal} {Journal Club for Condensed Matter Physics}\ }
  (\bibinfo {year} {2019}),\ 10.36471/JCCM\_September\_2019\_02}\BibitemShut
  {NoStop}%
\bibitem [{\citenamefont {Blatter}\ \emph {et~al.}(1994)\citenamefont
  {Blatter}, \citenamefont {Feigel'man}, \citenamefont {Geshkenbein},
  \citenamefont {Larkin},\ and\ \citenamefont {Vinokur}}]{RevModPhys.66.1125}%
  \BibitemOpen
  \bibfield  {author} {\bibinfo {author} {\bibfnamefont {G.}~\bibnamefont
  {Blatter}}, \bibinfo {author} {\bibfnamefont {M.~V.}\ \bibnamefont
  {Feigel'man}}, \bibinfo {author} {\bibfnamefont {V.~B.}\ \bibnamefont
  {Geshkenbein}}, \bibinfo {author} {\bibfnamefont {A.~I.}\ \bibnamefont
  {Larkin}}, \ and\ \bibinfo {author} {\bibfnamefont {V.~M.}\ \bibnamefont
  {Vinokur}},\ }\href {\doibase 10.1103/RevModPhys.66.1125} {\bibfield
  {journal} {\bibinfo  {journal} {Rev. Mod. Phys.}\ }\textbf {\bibinfo {volume}
  {66}},\ \bibinfo {pages} {1125} (\bibinfo {year} {1994})}\BibitemShut
  {NoStop}%
\bibitem [{\citenamefont {Brandt}(1995)}]{Brandt_1995}%
  \BibitemOpen
  \bibfield  {author} {\bibinfo {author} {\bibfnamefont {E.~H.}\ \bibnamefont
  {Brandt}},\ }\href {\doibase 10.1088/0034-4885/58/11/003} {\bibfield
  {journal} {\bibinfo  {journal} {Reports on Progress in Physics}\ }\textbf
  {\bibinfo {volume} {58}},\ \bibinfo {pages} {1465} (\bibinfo {year}
  {1995})}\BibitemShut {NoStop}%
\bibitem [{\citenamefont {Friedel}\ \emph {et~al.}(1963)\citenamefont
  {Friedel}, \citenamefont {De~Gennes},\ and\ \citenamefont
  {Matricon}}]{doi:10.1063/1.1753804}%
  \BibitemOpen
  \bibfield  {author} {\bibinfo {author} {\bibfnamefont {J.}~\bibnamefont
  {Friedel}}, \bibinfo {author} {\bibfnamefont {P.~G.}\ \bibnamefont
  {De~Gennes}}, \ and\ \bibinfo {author} {\bibfnamefont {J.}~\bibnamefont
  {Matricon}},\ }\href {\doibase 10.1063/1.1753804} {\bibfield  {journal}
  {\bibinfo  {journal} {Applied Physics Letters}\ }\textbf {\bibinfo {volume}
  {2}},\ \bibinfo {pages} {119} (\bibinfo {year} {1963})},\ \Eprint
  {http://arxiv.org/abs/https://doi.org/10.1063/1.1753804}
  {https://doi.org/10.1063/1.1753804} \BibitemShut {NoStop}%
\bibitem [{\citenamefont {DE~GENNES}\ and\ \citenamefont
  {MATRICON}(1964)}]{RevModPhys.36.45}%
  \BibitemOpen
  \bibfield  {author} {\bibinfo {author} {\bibfnamefont {P.~G.}\ \bibnamefont
  {DE~GENNES}}\ and\ \bibinfo {author} {\bibfnamefont {J.}~\bibnamefont
  {MATRICON}},\ }\href {\doibase 10.1103/RevModPhys.36.45} {\bibfield
  {journal} {\bibinfo  {journal} {Rev. Mod. Phys.}\ }\textbf {\bibinfo {volume}
  {36}},\ \bibinfo {pages} {45} (\bibinfo {year} {1964})}\BibitemShut {NoStop}%
\bibitem [{\citenamefont {Kogan}(1981)}]{PhysRevB.24.1572}%
  \BibitemOpen
  \bibfield  {author} {\bibinfo {author} {\bibfnamefont {V.~G.}\ \bibnamefont
  {Kogan}},\ }\href {\doibase 10.1103/PhysRevB.24.1572} {\bibfield  {journal}
  {\bibinfo  {journal} {Phys. Rev. B}\ }\textbf {\bibinfo {volume} {24}},\
  \bibinfo {pages} {1572} (\bibinfo {year} {1981})}\BibitemShut {NoStop}%
\bibitem [{\citenamefont {Buzdin}\ \emph {et~al.}(2013)\citenamefont {Buzdin},
  \citenamefont {Mel'nikov},\ and\ \citenamefont {Samokhvalov}}]{Buzdin2013}%
  \BibitemOpen
  \bibfield  {author} {\bibinfo {author} {\bibfnamefont {A.~I.}\ \bibnamefont
  {Buzdin}}, \bibinfo {author} {\bibfnamefont {A.~S.}\ \bibnamefont
  {Mel'nikov}}, \ and\ \bibinfo {author} {\bibfnamefont {A.~V.}\ \bibnamefont
  {Samokhvalov}},\ }\href {\doibase 10.1007/s10948-013-2206-4} {\bibfield
  {journal} {\bibinfo  {journal} {Journal of Superconductivity and Novel
  Magnetism}\ }\textbf {\bibinfo {volume} {26}},\ \bibinfo {pages} {2853}
  (\bibinfo {year} {2013})}\BibitemShut {NoStop}%
\bibitem [{\citenamefont {Pearl}(1964)}]{doi:10.1063/1.1754056}%
  \BibitemOpen
  \bibfield  {author} {\bibinfo {author} {\bibfnamefont {J.}~\bibnamefont
  {Pearl}},\ }\href {\doibase 10.1063/1.1754056} {\bibfield  {journal}
  {\bibinfo  {journal} {Applied Physics Letters}\ }\textbf {\bibinfo {volume}
  {5}},\ \bibinfo {pages} {65} (\bibinfo {year} {1964})},\ \Eprint
  {http://arxiv.org/abs/https://doi.org/10.1063/1.1754056}
  {https://doi.org/10.1063/1.1754056} \BibitemShut {NoStop}%
\bibitem [{\citenamefont {Brandt}(1993)}]{PhysRevB.48.6699}%
  \BibitemOpen
  \bibfield  {author} {\bibinfo {author} {\bibfnamefont {E.~H.}\ \bibnamefont
  {Brandt}},\ }\href {\doibase 10.1103/PhysRevB.48.6699} {\bibfield  {journal}
  {\bibinfo  {journal} {Phys. Rev. B}\ }\textbf {\bibinfo {volume} {48}},\
  \bibinfo {pages} {6699} (\bibinfo {year} {1993})}\BibitemShut {NoStop}%
\bibitem [{\citenamefont {Herrera}\ \emph {et~al.}(2017)\citenamefont
  {Herrera}, \citenamefont {Guillam\'on}, \citenamefont {Galvis}, \citenamefont
  {Correa}, \citenamefont {Fente}, \citenamefont {Vieira}, \citenamefont
  {Suderow}, \citenamefont {Martynovich},\ and\ \citenamefont
  {Kogan}}]{PhysRevB.96.184502}%
  \BibitemOpen
  \bibfield  {author} {\bibinfo {author} {\bibfnamefont {E.}~\bibnamefont
  {Herrera}}, \bibinfo {author} {\bibfnamefont {I.}~\bibnamefont
  {Guillam\'on}}, \bibinfo {author} {\bibfnamefont {J.~A.}\ \bibnamefont
  {Galvis}}, \bibinfo {author} {\bibfnamefont {A.}~\bibnamefont {Correa}},
  \bibinfo {author} {\bibfnamefont {A.}~\bibnamefont {Fente}}, \bibinfo
  {author} {\bibfnamefont {S.}~\bibnamefont {Vieira}}, \bibinfo {author}
  {\bibfnamefont {H.}~\bibnamefont {Suderow}}, \bibinfo {author} {\bibfnamefont
  {A.~Y.}\ \bibnamefont {Martynovich}}, \ and\ \bibinfo {author} {\bibfnamefont
  {V.~G.}\ \bibnamefont {Kogan}},\ }\href {\doibase 10.1103/PhysRevB.96.184502}
  {\bibfield  {journal} {\bibinfo  {journal} {Phys. Rev. B}\ }\textbf {\bibinfo
  {volume} {96}},\ \bibinfo {pages} {184502} (\bibinfo {year}
  {2017})}\BibitemShut {NoStop}%
\bibitem [{\citenamefont {Klatt}\ \emph {et~al.}(2020)\citenamefont {Klatt},
  \citenamefont {Kim},\ and\ \citenamefont {Torquato}}]{klatt2020cloaking}%
  \BibitemOpen
  \bibfield  {author} {\bibinfo {author} {\bibfnamefont {M.~A.}\ \bibnamefont
  {Klatt}}, \bibinfo {author} {\bibfnamefont {J.}~\bibnamefont {Kim}}, \ and\
  \bibinfo {author} {\bibfnamefont {S.}~\bibnamefont {Torquato}},\ }\href@noop
  {} {\enquote {\bibinfo {title} {Cloaking the underlying long-range order of
  randomly perturbed lattices},}\ } (\bibinfo {year} {2020}),\ \Eprint
  {http://arxiv.org/abs/2001.08161} {arXiv:2001.08161 [cond-mat.stat-mech]}
  \BibitemShut {NoStop}%
\bibitem [{\citenamefont {Guillam{\'o}n}\ \emph {et~al.}(2014)\citenamefont
  {Guillam{\'o}n}, \citenamefont {C{\'o}rdoba}, \citenamefont {Ses{\'e}},
  \citenamefont {De~Teresa}, \citenamefont {Ibarra}, \citenamefont {Vieira},\
  and\ \citenamefont {Suderow}}]{Guillamon2014}%
  \BibitemOpen
  \bibfield  {author} {\bibinfo {author} {\bibfnamefont {I.}~\bibnamefont
  {Guillam{\'o}n}}, \bibinfo {author} {\bibfnamefont {R.}~\bibnamefont
  {C{\'o}rdoba}}, \bibinfo {author} {\bibfnamefont {J.}~\bibnamefont
  {Ses{\'e}}}, \bibinfo {author} {\bibfnamefont {J.~M.}\ \bibnamefont
  {De~Teresa}}, \bibinfo {author} {\bibfnamefont {M.~R.}\ \bibnamefont
  {Ibarra}}, \bibinfo {author} {\bibfnamefont {S.}~\bibnamefont {Vieira}}, \
  and\ \bibinfo {author} {\bibfnamefont {H.}~\bibnamefont {Suderow}},\ }\href
  {https://doi.org/10.1038/nphys3132} {\bibfield  {journal} {\bibinfo
  {journal} {Nature Physics}\ }\textbf {\bibinfo {volume} {10}},\ \bibinfo
  {pages} {851 EP } (\bibinfo {year} {2014})}\BibitemShut {NoStop}%
\bibitem [{\citenamefont {Iavarone}\ \emph {et~al.}(2008)\citenamefont
  {Iavarone}, \citenamefont {Di~Capua}, \citenamefont {Karapetrov},
  \citenamefont {Koshelev}, \citenamefont {Rosenmann}, \citenamefont {Claus},
  \citenamefont {Malliakas}, \citenamefont {Kanatzidis}, \citenamefont
  {Nishizaki},\ and\ \citenamefont {Kobayashi}}]{PhysRevB.78.174518}%
  \BibitemOpen
  \bibfield  {author} {\bibinfo {author} {\bibfnamefont {M.}~\bibnamefont
  {Iavarone}}, \bibinfo {author} {\bibfnamefont {R.}~\bibnamefont {Di~Capua}},
  \bibinfo {author} {\bibfnamefont {G.}~\bibnamefont {Karapetrov}}, \bibinfo
  {author} {\bibfnamefont {A.~E.}\ \bibnamefont {Koshelev}}, \bibinfo {author}
  {\bibfnamefont {D.}~\bibnamefont {Rosenmann}}, \bibinfo {author}
  {\bibfnamefont {H.}~\bibnamefont {Claus}}, \bibinfo {author} {\bibfnamefont
  {C.~D.}\ \bibnamefont {Malliakas}}, \bibinfo {author} {\bibfnamefont {M.~G.}\
  \bibnamefont {Kanatzidis}}, \bibinfo {author} {\bibfnamefont
  {T.}~\bibnamefont {Nishizaki}}, \ and\ \bibinfo {author} {\bibfnamefont
  {N.}~\bibnamefont {Kobayashi}},\ }\href {\doibase 10.1103/PhysRevB.78.174518}
  {\bibfield  {journal} {\bibinfo  {journal} {Phys. Rev. B}\ }\textbf {\bibinfo
  {volume} {78}},\ \bibinfo {pages} {174518} (\bibinfo {year}
  {2008})}\BibitemShut {NoStop}%
\bibitem [{\citenamefont {Hanaguri}\ \emph {et~al.}(2012)\citenamefont
  {Hanaguri}, \citenamefont {Kitagawa}, \citenamefont {Matsubayashi},
  \citenamefont {Mazaki}, \citenamefont {Uwatoko},\ and\ \citenamefont
  {Takagi}}]{PhysRevB.85.214505}%
  \BibitemOpen
  \bibfield  {author} {\bibinfo {author} {\bibfnamefont {T.}~\bibnamefont
  {Hanaguri}}, \bibinfo {author} {\bibfnamefont {K.}~\bibnamefont {Kitagawa}},
  \bibinfo {author} {\bibfnamefont {K.}~\bibnamefont {Matsubayashi}}, \bibinfo
  {author} {\bibfnamefont {Y.}~\bibnamefont {Mazaki}}, \bibinfo {author}
  {\bibfnamefont {Y.}~\bibnamefont {Uwatoko}}, \ and\ \bibinfo {author}
  {\bibfnamefont {H.}~\bibnamefont {Takagi}},\ }\href {\doibase
  10.1103/PhysRevB.85.214505} {\bibfield  {journal} {\bibinfo  {journal} {Phys.
  Rev. B}\ }\textbf {\bibinfo {volume} {85}},\ \bibinfo {pages} {214505}
  (\bibinfo {year} {2012})}\BibitemShut {NoStop}%
\bibitem [{\citenamefont {Fente}\ \emph {et~al.}(2018)\citenamefont {Fente},
  \citenamefont {Meier}, \citenamefont {Kong}, \citenamefont {Kogan},
  \citenamefont {Bud'ko}, \citenamefont {Canfield}, \citenamefont
  {Guillam\'on},\ and\ \citenamefont {Suderow}}]{PhysRevB.97.134501}%
  \BibitemOpen
  \bibfield  {author} {\bibinfo {author} {\bibfnamefont {A.}~\bibnamefont
  {Fente}}, \bibinfo {author} {\bibfnamefont {W.~R.}\ \bibnamefont {Meier}},
  \bibinfo {author} {\bibfnamefont {T.}~\bibnamefont {Kong}}, \bibinfo {author}
  {\bibfnamefont {V.~G.}\ \bibnamefont {Kogan}}, \bibinfo {author}
  {\bibfnamefont {S.~L.}\ \bibnamefont {Bud'ko}}, \bibinfo {author}
  {\bibfnamefont {P.~C.}\ \bibnamefont {Canfield}}, \bibinfo {author}
  {\bibfnamefont {I.}~\bibnamefont {Guillam\'on}}, \ and\ \bibinfo {author}
  {\bibfnamefont {H.}~\bibnamefont {Suderow}},\ }\href {\doibase
  10.1103/PhysRevB.97.134501} {\bibfield  {journal} {\bibinfo  {journal} {Phys.
  Rev. B}\ }\textbf {\bibinfo {volume} {97}},\ \bibinfo {pages} {134501}
  (\bibinfo {year} {2018})}\BibitemShut {NoStop}%
\bibitem [{\citenamefont {Guillamón}\ \emph {et~al.}(2014)\citenamefont
  {Guillamón}, \citenamefont {Suderow}, \citenamefont {Kulkarni},
  \citenamefont {Vieira}, \citenamefont {Córdoba}, \citenamefont {Sesé},
  \citenamefont {Teresa}, \citenamefont {Ibarra}, \citenamefont {Shaw},\ and\
  \citenamefont {Bannerjee}}]{GUILLAMON201470}%
  \BibitemOpen
  \bibfield  {author} {\bibinfo {author} {\bibfnamefont {I.}~\bibnamefont
  {Guillamón}}, \bibinfo {author} {\bibfnamefont {H.}~\bibnamefont {Suderow}},
  \bibinfo {author} {\bibfnamefont {P.}~\bibnamefont {Kulkarni}}, \bibinfo
  {author} {\bibfnamefont {S.}~\bibnamefont {Vieira}}, \bibinfo {author}
  {\bibfnamefont {R.}~\bibnamefont {Córdoba}}, \bibinfo {author}
  {\bibfnamefont {J.}~\bibnamefont {Sesé}}, \bibinfo {author} {\bibfnamefont
  {J.~D.}\ \bibnamefont {Teresa}}, \bibinfo {author} {\bibfnamefont
  {M.}~\bibnamefont {Ibarra}}, \bibinfo {author} {\bibfnamefont
  {G.}~\bibnamefont {Shaw}}, \ and\ \bibinfo {author} {\bibfnamefont
  {S.}~\bibnamefont {Bannerjee}},\ }\href {\doibase
  https://doi.org/10.1016/j.physc.2014.04.031} {\bibfield  {journal} {\bibinfo
  {journal} {Physica C: Superconductivity and its Applications}\ }\textbf
  {\bibinfo {volume} {503}},\ \bibinfo {pages} {70 } (\bibinfo {year}
  {2014})}\BibitemShut {NoStop}%
\bibitem [{\citenamefont {Hess}\ \emph {et~al.}(1989)\citenamefont {Hess},
  \citenamefont {Robinson}, \citenamefont {Dynes}, \citenamefont {Valles},\
  and\ \citenamefont {Waszczak}}]{PhysRevLett.62.214}%
  \BibitemOpen
  \bibfield  {author} {\bibinfo {author} {\bibfnamefont {H.~F.}\ \bibnamefont
  {Hess}}, \bibinfo {author} {\bibfnamefont {R.~B.}\ \bibnamefont {Robinson}},
  \bibinfo {author} {\bibfnamefont {R.~C.}\ \bibnamefont {Dynes}}, \bibinfo
  {author} {\bibfnamefont {J.~M.}\ \bibnamefont {Valles}}, \ and\ \bibinfo
  {author} {\bibfnamefont {J.~V.}\ \bibnamefont {Waszczak}},\ }\href {\doibase
  10.1103/PhysRevLett.62.214} {\bibfield  {journal} {\bibinfo  {journal} {Phys.
  Rev. Lett.}\ }\textbf {\bibinfo {volume} {62}},\ \bibinfo {pages} {214}
  (\bibinfo {year} {1989})}\BibitemShut {NoStop}%
\bibitem [{\citenamefont {Guillam\'on}\ \emph {et~al.}(2008)\citenamefont
  {Guillam\'on}, \citenamefont {Suderow}, \citenamefont {Vieira}, \citenamefont
  {Cario}, \citenamefont {Diener},\ and\ \citenamefont
  {Rodi\`ere}}]{PhysRevLett.101.166407}%
  \BibitemOpen
  \bibfield  {author} {\bibinfo {author} {\bibfnamefont {I.}~\bibnamefont
  {Guillam\'on}}, \bibinfo {author} {\bibfnamefont {H.}~\bibnamefont
  {Suderow}}, \bibinfo {author} {\bibfnamefont {S.}~\bibnamefont {Vieira}},
  \bibinfo {author} {\bibfnamefont {L.}~\bibnamefont {Cario}}, \bibinfo
  {author} {\bibfnamefont {P.}~\bibnamefont {Diener}}, \ and\ \bibinfo {author}
  {\bibfnamefont {P.}~\bibnamefont {Rodi\`ere}},\ }\href {\doibase
  10.1103/PhysRevLett.101.166407} {\bibfield  {journal} {\bibinfo  {journal}
  {Phys. Rev. Lett.}\ }\textbf {\bibinfo {volume} {101}},\ \bibinfo {pages}
  {166407} (\bibinfo {year} {2008})}\BibitemShut {NoStop}%
\bibitem [{\citenamefont {Fischer}\ \emph {et~al.}(2007)\citenamefont
  {Fischer}, \citenamefont {Kugler}, \citenamefont {Maggio-Aprile},
  \citenamefont {Berthod},\ and\ \citenamefont {Renner}}]{RevModPhys.79.353}%
  \BibitemOpen
  \bibfield  {author} {\bibinfo {author} {\bibfnamefont {O.}~\bibnamefont
  {Fischer}}, \bibinfo {author} {\bibfnamefont {M.}~\bibnamefont {Kugler}},
  \bibinfo {author} {\bibfnamefont {I.}~\bibnamefont {Maggio-Aprile}}, \bibinfo
  {author} {\bibfnamefont {C.}~\bibnamefont {Berthod}}, \ and\ \bibinfo
  {author} {\bibfnamefont {C.}~\bibnamefont {Renner}},\ }\href {\doibase
  10.1103/RevModPhys.79.353} {\bibfield  {journal} {\bibinfo  {journal} {Rev.
  Mod. Phys.}\ }\textbf {\bibinfo {volume} {79}},\ \bibinfo {pages} {353}
  (\bibinfo {year} {2007})}\BibitemShut {NoStop}%
\bibitem [{\citenamefont {Suderow}\ \emph {et~al.}(2014)\citenamefont
  {Suderow}, \citenamefont {Guillam{\'{o}}n}, \citenamefont {Rodrigo},\ and\
  \citenamefont {Vieira}}]{Suderow_2014}%
  \BibitemOpen
  \bibfield  {author} {\bibinfo {author} {\bibfnamefont {H.}~\bibnamefont
  {Suderow}}, \bibinfo {author} {\bibfnamefont {I.}~\bibnamefont
  {Guillam{\'{o}}n}}, \bibinfo {author} {\bibfnamefont {J.~G.}\ \bibnamefont
  {Rodrigo}}, \ and\ \bibinfo {author} {\bibfnamefont {S.}~\bibnamefont
  {Vieira}},\ }\href {\doibase 10.1088/0953-2048/27/6/063001} {\bibfield
  {journal} {\bibinfo  {journal} {Superconductor Science and Technology}\
  }\textbf {\bibinfo {volume} {27}},\ \bibinfo {pages} {063001} (\bibinfo
  {year} {2014})}\BibitemShut {NoStop}%
\bibitem [{\citenamefont {Troyanovski}\ \emph {et~al.}(1999)\citenamefont
  {Troyanovski}, \citenamefont {Aarts},\ and\ \citenamefont
  {Kes}}]{Troyanovski1999}%
  \BibitemOpen
  \bibfield  {author} {\bibinfo {author} {\bibfnamefont {A.~M.}\ \bibnamefont
  {Troyanovski}}, \bibinfo {author} {\bibfnamefont {J.}~\bibnamefont {Aarts}},
  \ and\ \bibinfo {author} {\bibfnamefont {P.~H.}\ \bibnamefont {Kes}},\ }\href
  {\doibase 10.1038/21385} {\bibfield  {journal} {\bibinfo  {journal} {Nature}\
  }\textbf {\bibinfo {volume} {399}},\ \bibinfo {pages} {665} (\bibinfo {year}
  {1999})}\BibitemShut {NoStop}%
\bibitem [{\citenamefont {Suderow}\ \emph {et~al.}(2011)\citenamefont
  {Suderow}, \citenamefont {Guillamon},\ and\ \citenamefont
  {Vieira}}]{doi:10.1063/1.3567008}%
  \BibitemOpen
  \bibfield  {author} {\bibinfo {author} {\bibfnamefont {H.}~\bibnamefont
  {Suderow}}, \bibinfo {author} {\bibfnamefont {I.}~\bibnamefont {Guillamon}},
  \ and\ \bibinfo {author} {\bibfnamefont {S.}~\bibnamefont {Vieira}},\ }\href
  {\doibase 10.1063/1.3567008} {\bibfield  {journal} {\bibinfo  {journal}
  {Review of Scientific Instruments}\ }\textbf {\bibinfo {volume} {82}},\
  \bibinfo {pages} {033711} (\bibinfo {year} {2011})},\ \Eprint
  {http://arxiv.org/abs/https://doi.org/10.1063/1.3567008}
  {https://doi.org/10.1063/1.3567008} \BibitemShut {NoStop}%
\bibitem [{\citenamefont {Guillam{\'{o}}n}\ \emph {et~al.}(2008)\citenamefont
  {Guillam{\'{o}}n}, \citenamefont {Suderow}, \citenamefont {Vieira},
  \citenamefont {Fern{\'{a}}ndez-Pacheco}, \citenamefont {Ses{\'{e}}},
  \citenamefont {C{\'{o}}rdoba}, \citenamefont {Teresa},\ and\ \citenamefont
  {Ibarra}}]{Guillam_n_2008}%
  \BibitemOpen
  \bibfield  {author} {\bibinfo {author} {\bibfnamefont {I.}~\bibnamefont
  {Guillam{\'{o}}n}}, \bibinfo {author} {\bibfnamefont {H.}~\bibnamefont
  {Suderow}}, \bibinfo {author} {\bibfnamefont {S.}~\bibnamefont {Vieira}},
  \bibinfo {author} {\bibfnamefont {A.}~\bibnamefont
  {Fern{\'{a}}ndez-Pacheco}}, \bibinfo {author} {\bibfnamefont
  {J.}~\bibnamefont {Ses{\'{e}}}}, \bibinfo {author} {\bibfnamefont
  {R.}~\bibnamefont {C{\'{o}}rdoba}}, \bibinfo {author} {\bibfnamefont
  {J.~M.~D.}\ \bibnamefont {Teresa}}, \ and\ \bibinfo {author} {\bibfnamefont
  {M.~R.}\ \bibnamefont {Ibarra}},\ }\href {\doibase
  10.1088/1367-2630/10/9/093005} {\bibfield  {journal} {\bibinfo  {journal}
  {New Journal of Physics}\ }\textbf {\bibinfo {volume} {10}},\ \bibinfo
  {pages} {093005} (\bibinfo {year} {2008})}\BibitemShut {NoStop}%
\bibitem [{\citenamefont {C{\'o}rdoba}\ \emph {et~al.}(2013)\citenamefont
  {C{\'o}rdoba}, \citenamefont {Baturina}, \citenamefont {Ses{\'e}},
  \citenamefont {Yu~Mironov}, \citenamefont {De~Teresa}, \citenamefont
  {Ibarra}, \citenamefont {Nasimov}, \citenamefont {Gutakovskii}, \citenamefont
  {Latyshev}, \citenamefont {Guillam{\'o}n}, \citenamefont {Suderow},
  \citenamefont {Vieira}, \citenamefont {Baklanov}, \citenamefont {Palacios},\
  and\ \citenamefont {Vinokur}}]{Cordoba2013}%
  \BibitemOpen
  \bibfield  {author} {\bibinfo {author} {\bibfnamefont {R.}~\bibnamefont
  {C{\'o}rdoba}}, \bibinfo {author} {\bibfnamefont {T.~I.}\ \bibnamefont
  {Baturina}}, \bibinfo {author} {\bibfnamefont {J.}~\bibnamefont {Ses{\'e}}},
  \bibinfo {author} {\bibfnamefont {A.}~\bibnamefont {Yu~Mironov}}, \bibinfo
  {author} {\bibfnamefont {J.~M.}\ \bibnamefont {De~Teresa}}, \bibinfo {author}
  {\bibfnamefont {M.~R.}\ \bibnamefont {Ibarra}}, \bibinfo {author}
  {\bibfnamefont {D.~A.}\ \bibnamefont {Nasimov}}, \bibinfo {author}
  {\bibfnamefont {A.~K.}\ \bibnamefont {Gutakovskii}}, \bibinfo {author}
  {\bibfnamefont {A.~V.}\ \bibnamefont {Latyshev}}, \bibinfo {author}
  {\bibfnamefont {I.}~\bibnamefont {Guillam{\'o}n}}, \bibinfo {author}
  {\bibfnamefont {H.}~\bibnamefont {Suderow}}, \bibinfo {author} {\bibfnamefont
  {S.}~\bibnamefont {Vieira}}, \bibinfo {author} {\bibfnamefont {M.~R.}\
  \bibnamefont {Baklanov}}, \bibinfo {author} {\bibfnamefont {J.~J.}\
  \bibnamefont {Palacios}}, \ and\ \bibinfo {author} {\bibfnamefont {V.~M.}\
  \bibnamefont {Vinokur}},\ }\href {\doibase 10.1038/ncomms2437} {\bibfield
  {journal} {\bibinfo  {journal} {Nature Communications}\ }\textbf {\bibinfo
  {volume} {4}},\ \bibinfo {pages} {1437} (\bibinfo {year} {2013})}\BibitemShut
  {NoStop}%
\bibitem [{\citenamefont {Klein}\ \emph {et~al.}(2001)\citenamefont {Klein},
  \citenamefont {Joumard}, \citenamefont {Blanchard}, \citenamefont {Marcus},
  \citenamefont {Cubitt}, \citenamefont {Giamarchi},\ and\ \citenamefont
  {Le~Doussal}}]{Klein2001}%
  \BibitemOpen
  \bibfield  {author} {\bibinfo {author} {\bibfnamefont {T.}~\bibnamefont
  {Klein}}, \bibinfo {author} {\bibfnamefont {I.}~\bibnamefont {Joumard}},
  \bibinfo {author} {\bibfnamefont {S.}~\bibnamefont {Blanchard}}, \bibinfo
  {author} {\bibfnamefont {J.}~\bibnamefont {Marcus}}, \bibinfo {author}
  {\bibfnamefont {R.}~\bibnamefont {Cubitt}}, \bibinfo {author} {\bibfnamefont
  {T.}~\bibnamefont {Giamarchi}}, \ and\ \bibinfo {author} {\bibfnamefont
  {P.}~\bibnamefont {Le~Doussal}},\ }\href {\doibase 10.1038/35096534}
  {\bibfield  {journal} {\bibinfo  {journal} {Nature}\ }\textbf {\bibinfo
  {volume} {413}},\ \bibinfo {pages} {404} (\bibinfo {year}
  {2001})}\BibitemShut {NoStop}%
\bibitem [{\citenamefont {Giamarchi}\ and\ \citenamefont
  {Le~Doussal}(1994)}]{PhysRevLett.72.1530}%
  \BibitemOpen
  \bibfield  {author} {\bibinfo {author} {\bibfnamefont {T.}~\bibnamefont
  {Giamarchi}}\ and\ \bibinfo {author} {\bibfnamefont {P.}~\bibnamefont
  {Le~Doussal}},\ }\href {\doibase 10.1103/PhysRevLett.72.1530} {\bibfield
  {journal} {\bibinfo  {journal} {Phys. Rev. Lett.}\ }\textbf {\bibinfo
  {volume} {72}},\ \bibinfo {pages} {1530} (\bibinfo {year}
  {1994})}\BibitemShut {NoStop}%
\bibitem [{\citenamefont {Giamarchi}\ and\ \citenamefont
  {Le~Doussal}(1995)}]{PhysRevB.52.1242}%
  \BibitemOpen
  \bibfield  {author} {\bibinfo {author} {\bibfnamefont {T.}~\bibnamefont
  {Giamarchi}}\ and\ \bibinfo {author} {\bibfnamefont {P.}~\bibnamefont
  {Le~Doussal}},\ }\href {\doibase 10.1103/PhysRevB.52.1242} {\bibfield
  {journal} {\bibinfo  {journal} {Phys. Rev. B}\ }\textbf {\bibinfo {volume}
  {52}},\ \bibinfo {pages} {1242} (\bibinfo {year} {1995})}\BibitemShut
  {NoStop}%
\bibitem [{\citenamefont {Fisher}\ \emph {et~al.}(1991)\citenamefont {Fisher},
  \citenamefont {Fisher},\ and\ \citenamefont {Huse}}]{PhysRevB.43.130}%
  \BibitemOpen
  \bibfield  {author} {\bibinfo {author} {\bibfnamefont {D.~S.}\ \bibnamefont
  {Fisher}}, \bibinfo {author} {\bibfnamefont {M.~P.~A.}\ \bibnamefont
  {Fisher}}, \ and\ \bibinfo {author} {\bibfnamefont {D.~A.}\ \bibnamefont
  {Huse}},\ }\href {\doibase 10.1103/PhysRevB.43.130} {\bibfield  {journal}
  {\bibinfo  {journal} {Phys. Rev. B}\ }\textbf {\bibinfo {volume} {43}},\
  \bibinfo {pages} {130} (\bibinfo {year} {1991})}\BibitemShut {NoStop}%
\bibitem [{\citenamefont {Kierfeld}\ \emph {et~al.}(1997)\citenamefont
  {Kierfeld}, \citenamefont {Nattermann},\ and\ \citenamefont
  {Hwa}}]{PhysRevB.55.626}%
  \BibitemOpen
  \bibfield  {author} {\bibinfo {author} {\bibfnamefont {J.}~\bibnamefont
  {Kierfeld}}, \bibinfo {author} {\bibfnamefont {T.}~\bibnamefont
  {Nattermann}}, \ and\ \bibinfo {author} {\bibfnamefont {T.}~\bibnamefont
  {Hwa}},\ }\href {\doibase 10.1103/PhysRevB.55.626} {\bibfield  {journal}
  {\bibinfo  {journal} {Phys. Rev. B}\ }\textbf {\bibinfo {volume} {55}},\
  \bibinfo {pages} {626} (\bibinfo {year} {1997})}\BibitemShut {NoStop}%
\bibitem [{\citenamefont {Singh}\ \emph {et~al.}(2018)\citenamefont {Singh},
  \citenamefont {Bristow}, \citenamefont {Meier}, \citenamefont {Taylor},
  \citenamefont {Blundell}, \citenamefont {Canfield},\ and\ \citenamefont
  {Coldea}}]{PhysRevMaterials.2.074802}%
  \BibitemOpen
  \bibfield  {author} {\bibinfo {author} {\bibfnamefont {S.~J.}\ \bibnamefont
  {Singh}}, \bibinfo {author} {\bibfnamefont {M.}~\bibnamefont {Bristow}},
  \bibinfo {author} {\bibfnamefont {W.~R.}\ \bibnamefont {Meier}}, \bibinfo
  {author} {\bibfnamefont {P.}~\bibnamefont {Taylor}}, \bibinfo {author}
  {\bibfnamefont {S.~J.}\ \bibnamefont {Blundell}}, \bibinfo {author}
  {\bibfnamefont {P.~C.}\ \bibnamefont {Canfield}}, \ and\ \bibinfo {author}
  {\bibfnamefont {A.~I.}\ \bibnamefont {Coldea}},\ }\href {\doibase
  10.1103/PhysRevMaterials.2.074802} {\bibfield  {journal} {\bibinfo  {journal}
  {Phys. Rev. Materials}\ }\textbf {\bibinfo {volume} {2}},\ \bibinfo {pages}
  {074802} (\bibinfo {year} {2018})}\BibitemShut {NoStop}%
\bibitem [{\citenamefont {Pyon}\ \emph {et~al.}(2019)\citenamefont {Pyon},
  \citenamefont {Takahashi}, \citenamefont {Veshchunov}, \citenamefont
  {Tamegai}, \citenamefont {Ishida}, \citenamefont {Iyo}, \citenamefont
  {Eisaki}, \citenamefont {Imai}, \citenamefont {Abe}, \citenamefont
  {Terashima},\ and\ \citenamefont {Ichinose}}]{PhysRevB.99.104506}%
  \BibitemOpen
  \bibfield  {author} {\bibinfo {author} {\bibfnamefont {S.}~\bibnamefont
  {Pyon}}, \bibinfo {author} {\bibfnamefont {A.}~\bibnamefont {Takahashi}},
  \bibinfo {author} {\bibfnamefont {I.}~\bibnamefont {Veshchunov}}, \bibinfo
  {author} {\bibfnamefont {T.}~\bibnamefont {Tamegai}}, \bibinfo {author}
  {\bibfnamefont {S.}~\bibnamefont {Ishida}}, \bibinfo {author} {\bibfnamefont
  {A.}~\bibnamefont {Iyo}}, \bibinfo {author} {\bibfnamefont {H.}~\bibnamefont
  {Eisaki}}, \bibinfo {author} {\bibfnamefont {M.}~\bibnamefont {Imai}},
  \bibinfo {author} {\bibfnamefont {H.}~\bibnamefont {Abe}}, \bibinfo {author}
  {\bibfnamefont {T.}~\bibnamefont {Terashima}}, \ and\ \bibinfo {author}
  {\bibfnamefont {A.}~\bibnamefont {Ichinose}},\ }\href {\doibase
  10.1103/PhysRevB.99.104506} {\bibfield  {journal} {\bibinfo  {journal} {Phys.
  Rev. B}\ }\textbf {\bibinfo {volume} {99}},\ \bibinfo {pages} {104506}
  (\bibinfo {year} {2019})}\BibitemShut {NoStop}%
\bibitem [{\citenamefont {Haberkorn}\ \emph {et~al.}(2019)\citenamefont
  {Haberkorn}, \citenamefont {Xu}, \citenamefont {Meier}, \citenamefont
  {Schmidt}, \citenamefont {Bud'ko},\ and\ \citenamefont
  {Canfield}}]{PhysRevB.100.064524}%
  \BibitemOpen
  \bibfield  {author} {\bibinfo {author} {\bibfnamefont {N.}~\bibnamefont
  {Haberkorn}}, \bibinfo {author} {\bibfnamefont {M.}~\bibnamefont {Xu}},
  \bibinfo {author} {\bibfnamefont {W.~R.}\ \bibnamefont {Meier}}, \bibinfo
  {author} {\bibfnamefont {J.}~\bibnamefont {Schmidt}}, \bibinfo {author}
  {\bibfnamefont {S.~L.}\ \bibnamefont {Bud'ko}}, \ and\ \bibinfo {author}
  {\bibfnamefont {P.~C.}\ \bibnamefont {Canfield}},\ }\href {\doibase
  10.1103/PhysRevB.100.064524} {\bibfield  {journal} {\bibinfo  {journal}
  {Phys. Rev. B}\ }\textbf {\bibinfo {volume} {100}},\ \bibinfo {pages}
  {064524} (\bibinfo {year} {2019})}\BibitemShut {NoStop}%
\bibitem [{\citenamefont {Ishida}\ \emph {et~al.}(2019)\citenamefont {Ishida},
  \citenamefont {Iyo}, \citenamefont {Ogino}, \citenamefont {Eisaki},
  \citenamefont {Takeshita}, \citenamefont {Kawashima}, \citenamefont
  {Yanagisawa}, \citenamefont {Kobayashi}, \citenamefont {Kimoto},
  \citenamefont {Abe}, \citenamefont {Imai}, \citenamefont {Shimoyama},\ and\
  \citenamefont {Eisterer}}]{Ishida2019}%
  \BibitemOpen
  \bibfield  {author} {\bibinfo {author} {\bibfnamefont {S.}~\bibnamefont
  {Ishida}}, \bibinfo {author} {\bibfnamefont {A.}~\bibnamefont {Iyo}},
  \bibinfo {author} {\bibfnamefont {H.}~\bibnamefont {Ogino}}, \bibinfo
  {author} {\bibfnamefont {H.}~\bibnamefont {Eisaki}}, \bibinfo {author}
  {\bibfnamefont {N.}~\bibnamefont {Takeshita}}, \bibinfo {author}
  {\bibfnamefont {K.}~\bibnamefont {Kawashima}}, \bibinfo {author}
  {\bibfnamefont {K.}~\bibnamefont {Yanagisawa}}, \bibinfo {author}
  {\bibfnamefont {Y.}~\bibnamefont {Kobayashi}}, \bibinfo {author}
  {\bibfnamefont {K.}~\bibnamefont {Kimoto}}, \bibinfo {author} {\bibfnamefont
  {H.}~\bibnamefont {Abe}}, \bibinfo {author} {\bibfnamefont {M.}~\bibnamefont
  {Imai}}, \bibinfo {author} {\bibfnamefont {J.-i.}\ \bibnamefont {Shimoyama}},
  \ and\ \bibinfo {author} {\bibfnamefont {M.}~\bibnamefont {Eisterer}},\
  }\href {\doibase 10.1038/s41535-019-0165-0} {\bibfield  {journal} {\bibinfo
  {journal} {npj Quantum Materials}\ }\textbf {\bibinfo {volume} {4}},\
  \bibinfo {pages} {27} (\bibinfo {year} {2019})}\BibitemShut {NoStop}%
\bibitem [{\citenamefont {Kogan}\ \emph {et~al.}(1997)\citenamefont {Kogan},
  \citenamefont {Bullock}, \citenamefont {Harmon}, \citenamefont
  {Miranovic-acute}, \citenamefont {Dobrosavljevic-acute Grujic-acute},
  \citenamefont {Gammel},\ and\ \citenamefont {Bishop}}]{PhysRevB.55.R8693}%
  \BibitemOpen
  \bibfield  {author} {\bibinfo {author} {\bibfnamefont {V.~G.}\ \bibnamefont
  {Kogan}}, \bibinfo {author} {\bibfnamefont {M.}~\bibnamefont {Bullock}},
  \bibinfo {author} {\bibfnamefont {B.}~\bibnamefont {Harmon}}, \bibinfo
  {author} {\bibfnamefont {P.}~\bibnamefont {Miranovic-acute}}, \bibinfo
  {author} {\bibfnamefont {L.}~\bibnamefont {Dobrosavljevic-acute
  Grujic-acute}}, \bibinfo {author} {\bibfnamefont {P.~L.}\ \bibnamefont
  {Gammel}}, \ and\ \bibinfo {author} {\bibfnamefont {D.~J.}\ \bibnamefont
  {Bishop}},\ }\href {\doibase 10.1103/PhysRevB.55.R8693} {\bibfield  {journal}
  {\bibinfo  {journal} {Phys. Rev. B}\ }\textbf {\bibinfo {volume} {55}},\
  \bibinfo {pages} {R8693} (\bibinfo {year} {1997})}\BibitemShut {NoStop}%
\bibitem [{\citenamefont {Zhang}\ \emph {et~al.}(2019)\citenamefont {Zhang},
  \citenamefont {Yin}, \citenamefont {Dai}, \citenamefont {Zheng},
  \citenamefont {Chang}, \citenamefont {Belopolski}, \citenamefont {Wang},
  \citenamefont {Lin}, \citenamefont {Wang}, \citenamefont {Jin},\ and\
  \citenamefont {Hasan}}]{PhysRevB.99.161103}%
  \BibitemOpen
  \bibfield  {author} {\bibinfo {author} {\bibfnamefont {S.~S.}\ \bibnamefont
  {Zhang}}, \bibinfo {author} {\bibfnamefont {J.-X.}\ \bibnamefont {Yin}},
  \bibinfo {author} {\bibfnamefont {G.}~\bibnamefont {Dai}}, \bibinfo {author}
  {\bibfnamefont {H.}~\bibnamefont {Zheng}}, \bibinfo {author} {\bibfnamefont
  {G.}~\bibnamefont {Chang}}, \bibinfo {author} {\bibfnamefont
  {I.}~\bibnamefont {Belopolski}}, \bibinfo {author} {\bibfnamefont
  {X.}~\bibnamefont {Wang}}, \bibinfo {author} {\bibfnamefont {H.}~\bibnamefont
  {Lin}}, \bibinfo {author} {\bibfnamefont {Z.}~\bibnamefont {Wang}}, \bibinfo
  {author} {\bibfnamefont {C.}~\bibnamefont {Jin}}, \ and\ \bibinfo {author}
  {\bibfnamefont {M.~Z.}\ \bibnamefont {Hasan}},\ }\href {\doibase
  10.1103/PhysRevB.99.161103} {\bibfield  {journal} {\bibinfo  {journal} {Phys.
  Rev. B}\ }\textbf {\bibinfo {volume} {99}},\ \bibinfo {pages} {161103}
  (\bibinfo {year} {2019})}\BibitemShut {NoStop}%
\bibitem [{\citenamefont {Inosov}\ \emph {et~al.}(2010)\citenamefont {Inosov},
  \citenamefont {White}, \citenamefont {Evtushinsky}, \citenamefont {Morozov},
  \citenamefont {Cameron}, \citenamefont {Stockert}, \citenamefont
  {Zabolotnyy}, \citenamefont {Kim}, \citenamefont {Kordyuk}, \citenamefont
  {Borisenko}, \citenamefont {Forgan}, \citenamefont {Klingeler}, \citenamefont
  {Park}, \citenamefont {Wurmehl}, \citenamefont {Vasiliev}, \citenamefont
  {Behr}, \citenamefont {Dewhurst},\ and\ \citenamefont
  {Hinkov}}]{PhysRevLett.104.187001}%
  \BibitemOpen
  \bibfield  {author} {\bibinfo {author} {\bibfnamefont {D.~S.}\ \bibnamefont
  {Inosov}}, \bibinfo {author} {\bibfnamefont {J.~S.}\ \bibnamefont {White}},
  \bibinfo {author} {\bibfnamefont {D.~V.}\ \bibnamefont {Evtushinsky}},
  \bibinfo {author} {\bibfnamefont {I.~V.}\ \bibnamefont {Morozov}}, \bibinfo
  {author} {\bibfnamefont {A.}~\bibnamefont {Cameron}}, \bibinfo {author}
  {\bibfnamefont {U.}~\bibnamefont {Stockert}}, \bibinfo {author}
  {\bibfnamefont {V.~B.}\ \bibnamefont {Zabolotnyy}}, \bibinfo {author}
  {\bibfnamefont {T.~K.}\ \bibnamefont {Kim}}, \bibinfo {author} {\bibfnamefont
  {A.~A.}\ \bibnamefont {Kordyuk}}, \bibinfo {author} {\bibfnamefont {S.~V.}\
  \bibnamefont {Borisenko}}, \bibinfo {author} {\bibfnamefont {E.~M.}\
  \bibnamefont {Forgan}}, \bibinfo {author} {\bibfnamefont {R.}~\bibnamefont
  {Klingeler}}, \bibinfo {author} {\bibfnamefont {J.~T.}\ \bibnamefont {Park}},
  \bibinfo {author} {\bibfnamefont {S.}~\bibnamefont {Wurmehl}}, \bibinfo
  {author} {\bibfnamefont {A.~N.}\ \bibnamefont {Vasiliev}}, \bibinfo {author}
  {\bibfnamefont {G.}~\bibnamefont {Behr}}, \bibinfo {author} {\bibfnamefont
  {C.~D.}\ \bibnamefont {Dewhurst}}, \ and\ \bibinfo {author} {\bibfnamefont
  {V.}~\bibnamefont {Hinkov}},\ }\href {\doibase
  10.1103/PhysRevLett.104.187001} {\bibfield  {journal} {\bibinfo  {journal}
  {Phys. Rev. Lett.}\ }\textbf {\bibinfo {volume} {104}},\ \bibinfo {pages}
  {187001} (\bibinfo {year} {2010})}\BibitemShut {NoStop}%
\bibitem [{\citenamefont {Rumi}\ \emph {et~al.}(2019)\citenamefont {Rumi},
  \citenamefont {Arag\'on~S\'anchez}, \citenamefont {El\'{\i}as}, \citenamefont
  {Cort\'es~Maldonado}, \citenamefont {Puig}, \citenamefont {Cejas~Bolecek},
  \citenamefont {Nieva}, \citenamefont {Konczykowski}, \citenamefont {Fasano},\
  and\ \citenamefont {Kolton}}]{PhysRevResearch.1.033057}%
  \BibitemOpen
  \bibfield  {author} {\bibinfo {author} {\bibfnamefont {G.}~\bibnamefont
  {Rumi}}, \bibinfo {author} {\bibfnamefont {J.}~\bibnamefont
  {Arag\'on~S\'anchez}}, \bibinfo {author} {\bibfnamefont {F.}~\bibnamefont
  {El\'{\i}as}}, \bibinfo {author} {\bibfnamefont {R.}~\bibnamefont
  {Cort\'es~Maldonado}}, \bibinfo {author} {\bibfnamefont {J.}~\bibnamefont
  {Puig}}, \bibinfo {author} {\bibfnamefont {N.~R.}\ \bibnamefont
  {Cejas~Bolecek}}, \bibinfo {author} {\bibfnamefont {G.}~\bibnamefont
  {Nieva}}, \bibinfo {author} {\bibfnamefont {M.}~\bibnamefont {Konczykowski}},
  \bibinfo {author} {\bibfnamefont {Y.}~\bibnamefont {Fasano}}, \ and\ \bibinfo
  {author} {\bibfnamefont {A.~B.}\ \bibnamefont {Kolton}},\ }\href {\doibase
  10.1103/PhysRevResearch.1.033057} {\bibfield  {journal} {\bibinfo  {journal}
  {Phys. Rev. Research}\ }\textbf {\bibinfo {volume} {1}},\ \bibinfo {pages}
  {033057} (\bibinfo {year} {2019})}\BibitemShut {NoStop}%
\bibitem [{\citenamefont {Prozorov}\ \emph {et~al.}(2008)\citenamefont
  {Prozorov}, \citenamefont {Fidler}, \citenamefont {Hoberg},\ and\
  \citenamefont {Canfield}}]{Prozorov2008}%
  \BibitemOpen
  \bibfield  {author} {\bibinfo {author} {\bibfnamefont {R.}~\bibnamefont
  {Prozorov}}, \bibinfo {author} {\bibfnamefont {A.~F.}\ \bibnamefont
  {Fidler}}, \bibinfo {author} {\bibfnamefont {J.~R.}\ \bibnamefont {Hoberg}},
  \ and\ \bibinfo {author} {\bibfnamefont {P.~C.}\ \bibnamefont {Canfield}},\
  }\href {\doibase 10.1038/nphys888} {\bibfield  {journal} {\bibinfo  {journal}
  {Nature Physics}\ }\textbf {\bibinfo {volume} {4}},\ \bibinfo {pages} {327}
  (\bibinfo {year} {2008})}\BibitemShut {NoStop}%
\bibitem [{\citenamefont {Llorens}\ \emph {et~al.}(2019)\citenamefont
  {Llorens}, \citenamefont {Embon}, \citenamefont {Correa}, \citenamefont
  {González}, \citenamefont {Herrera}, \citenamefont {Guillamón},
  \citenamefont {Luccas}, \citenamefont {Azpeitia}, \citenamefont {Mompeán},
  \citenamefont {García-Hernández}, \citenamefont {Munuera}, \citenamefont
  {Sánchez}, \citenamefont {Fasano}, \citenamefont {Milosevic}, \citenamefont
  {Suderow},\ and\ \citenamefont {Anahory}}]{llorens2019observation}%
  \BibitemOpen
  \bibfield  {author} {\bibinfo {author} {\bibfnamefont {J.~B.}\ \bibnamefont
  {Llorens}}, \bibinfo {author} {\bibfnamefont {L.}~\bibnamefont {Embon}},
  \bibinfo {author} {\bibfnamefont {A.}~\bibnamefont {Correa}}, \bibinfo
  {author} {\bibfnamefont {J.~D.}\ \bibnamefont {González}}, \bibinfo {author}
  {\bibfnamefont {E.}~\bibnamefont {Herrera}}, \bibinfo {author} {\bibfnamefont
  {I.}~\bibnamefont {Guillamón}}, \bibinfo {author} {\bibfnamefont {R.~F.}\
  \bibnamefont {Luccas}}, \bibinfo {author} {\bibfnamefont {J.}~\bibnamefont
  {Azpeitia}}, \bibinfo {author} {\bibfnamefont {F.~J.}\ \bibnamefont
  {Mompeán}}, \bibinfo {author} {\bibfnamefont {M.}~\bibnamefont
  {García-Hernández}}, \bibinfo {author} {\bibfnamefont {C.}~\bibnamefont
  {Munuera}}, \bibinfo {author} {\bibfnamefont {J.~A.}\ \bibnamefont
  {Sánchez}}, \bibinfo {author} {\bibfnamefont {Y.}~\bibnamefont {Fasano}},
  \bibinfo {author} {\bibfnamefont {M.~V.}\ \bibnamefont {Milosevic}}, \bibinfo
  {author} {\bibfnamefont {H.}~\bibnamefont {Suderow}}, \ and\ \bibinfo
  {author} {\bibfnamefont {Y.}~\bibnamefont {Anahory}},\ }\href@noop {}
  {\enquote {\bibinfo {title} {Observation of a gel of quantum vortices in a
  superconductor at very low magnetic fields},}\ } (\bibinfo {year} {2019}),\
  \Eprint {http://arxiv.org/abs/1904.10999} {arXiv:1904.10999
  [cond-mat.supr-con]} \BibitemShut {NoStop}%
\bibitem [{\citenamefont {Cubitt}\ \emph {et~al.}(1993)\citenamefont {Cubitt},
  \citenamefont {Forgan}, \citenamefont {Yang}, \citenamefont {Lee},
  \citenamefont {Paul}, \citenamefont {Mook}, \citenamefont {Yethiraj},
  \citenamefont {Kes}, \citenamefont {Li}, \citenamefont {Menovsky},
  \citenamefont {Tarnawski},\ and\ \citenamefont {Mortensen}}]{Cubitt1993}%
  \BibitemOpen
  \bibfield  {author} {\bibinfo {author} {\bibfnamefont {R.}~\bibnamefont
  {Cubitt}}, \bibinfo {author} {\bibfnamefont {E.~M.}\ \bibnamefont {Forgan}},
  \bibinfo {author} {\bibfnamefont {G.}~\bibnamefont {Yang}}, \bibinfo {author}
  {\bibfnamefont {S.~L.}\ \bibnamefont {Lee}}, \bibinfo {author} {\bibfnamefont
  {D.~M.}\ \bibnamefont {Paul}}, \bibinfo {author} {\bibfnamefont {H.~A.}\
  \bibnamefont {Mook}}, \bibinfo {author} {\bibfnamefont {M.}~\bibnamefont
  {Yethiraj}}, \bibinfo {author} {\bibfnamefont {P.~H.}\ \bibnamefont {Kes}},
  \bibinfo {author} {\bibfnamefont {T.~W.}\ \bibnamefont {Li}}, \bibinfo
  {author} {\bibfnamefont {A.~A.}\ \bibnamefont {Menovsky}}, \bibinfo {author}
  {\bibfnamefont {Z.}~\bibnamefont {Tarnawski}}, \ and\ \bibinfo {author}
  {\bibfnamefont {K.}~\bibnamefont {Mortensen}},\ }\href {\doibase
  10.1038/365407a0} {\bibfield  {journal} {\bibinfo  {journal} {Nature}\
  }\textbf {\bibinfo {volume} {365}},\ \bibinfo {pages} {407} (\bibinfo {year}
  {1993})}\BibitemShut {NoStop}%
\bibitem [{\citenamefont {Zeldov}\ \emph {et~al.}(1995)\citenamefont {Zeldov},
  \citenamefont {Majer}, \citenamefont {Konczykowski}, \citenamefont
  {Geshkenbein}, \citenamefont {Vinokur},\ and\ \citenamefont
  {Shtrikman}}]{Zeldov1995}%
  \BibitemOpen
  \bibfield  {author} {\bibinfo {author} {\bibfnamefont {E.}~\bibnamefont
  {Zeldov}}, \bibinfo {author} {\bibfnamefont {D.}~\bibnamefont {Majer}},
  \bibinfo {author} {\bibfnamefont {M.}~\bibnamefont {Konczykowski}}, \bibinfo
  {author} {\bibfnamefont {V.~B.}\ \bibnamefont {Geshkenbein}}, \bibinfo
  {author} {\bibfnamefont {V.~M.}\ \bibnamefont {Vinokur}}, \ and\ \bibinfo
  {author} {\bibfnamefont {H.}~\bibnamefont {Shtrikman}},\ }\href {\doibase
  10.1038/375373a0} {\bibfield  {journal} {\bibinfo  {journal} {Nature}\
  }\textbf {\bibinfo {volume} {375}},\ \bibinfo {pages} {373} (\bibinfo {year}
  {1995})}\BibitemShut {NoStop}%
\bibitem [{\citenamefont {Marchevsky}\ \emph {et~al.}(1997)\citenamefont
  {Marchevsky}, \citenamefont {Kes},\ and\ \citenamefont
  {Aarts}}]{MARCHEVSKY19972083}%
  \BibitemOpen
  \bibfield  {author} {\bibinfo {author} {\bibfnamefont {M.}~\bibnamefont
  {Marchevsky}}, \bibinfo {author} {\bibfnamefont {P.}~\bibnamefont {Kes}}, \
  and\ \bibinfo {author} {\bibfnamefont {J.}~\bibnamefont {Aarts}},\ }\href
  {\doibase https://doi.org/10.1016/S0921-4534(97)01147-7} {\bibfield
  {journal} {\bibinfo  {journal} {Physica C: Superconductivity and its
  Applications}\ }\textbf {\bibinfo {volume} {282-287}},\ \bibinfo {pages}
  {2083 } (\bibinfo {year} {1997})}\BibitemShut {NoStop}%
\bibitem [{\citenamefont {Serrano}\ \emph {et~al.}(2016)\citenamefont
  {Serrano}, \citenamefont {Ses\'e}, \citenamefont {Guillam\'on}, \citenamefont
  {Suderow}, \citenamefont {Vieira}, \citenamefont {Ibarra},\ and\
  \citenamefont {De~Teresa}}]{PinningWbasedStrong}%
  \BibitemOpen
  \bibfield  {author} {\bibinfo {author} {\bibfnamefont {I.~G.}\ \bibnamefont
  {Serrano}}, \bibinfo {author} {\bibfnamefont {J.}~\bibnamefont {Ses\'e}},
  \bibinfo {author} {\bibfnamefont {I.}~\bibnamefont {Guillam\'on}}, \bibinfo
  {author} {\bibfnamefont {H.}~\bibnamefont {Suderow}}, \bibinfo {author}
  {\bibfnamefont {S.}~\bibnamefont {Vieira}}, \bibinfo {author} {\bibfnamefont
  {M.}~\bibnamefont {Ibarra}}, \ and\ \bibinfo {author} {\bibfnamefont {J.~M.}\
  \bibnamefont {De~Teresa}},\ }\href {\doibase doi:10.3762/bjnano.7.162}
  {\bibfield  {journal} {\bibinfo  {journal} {Beilstein J. Nanotechnol.}\ ,\
  \bibinfo {pages} {1698}} (\bibinfo {year} {2016})}\BibitemShut {NoStop}%
\end{thebibliography}

%

\end{document}